\documentclass[prd,twocolumn,a4paper,superscriptaddress]{revtex4}
\usepackage{graphicx}

\begin{document}

\newcommand{\be}{\begin{equation}}
\newcommand{\ee}{\end{equation}}
\newcommand{\bq}{\begin{eqnarray}}
\newcommand{\eq}{\end{eqnarray}}
\newcommand{\bsq}{\begin{subequations}}
\newcommand{\esq}{\end{subequations}}
\newcommand{\bc}{\begin{center}}
\newcommand{\ec}{\end{center}}

\newcommand {\R}{{\mathcal R}}
\newcommand{\al}{\alpha}

\title{Topological defects: A problem for cyclic universes?}

\author{P.P. Avelino}
\email[Electronic address: ]{ppavelin@fc.up.pt}
\affiliation{Centro de F\'\i sica do Porto e Departamento de F\'\i sica da
Faculdade de Ci\^encias da Universidade do
Porto, Rua do Campo Alegre 687, 4169-007, Porto, Portugal}
\author{C.J.A.P. Martins}
\email[Electronic address: ]{C.J.A.P.Martins@damtp.cam.ac.uk}
\affiliation{Centro de Astrof\'{\i}sica da Universidade do Porto, R. das
Estrelas s/n, 4150-762 Porto, Portugal}
\affiliation{Department of Applied Mathematics and Theoretical Physics,
Centre for Mathematical Sciences,\\ University of Cambridge,
Wilberforce Road, Cambridge CB3 0WA, United Kingdom}
\affiliation{Institut d'Astrophysique de Paris, 98 bis Boulevard Arago,
75014 Paris, France}
\author{C. Santos}
\email[Electronic address: ]{cssilva@fc.up.pt}
\affiliation{Centro de F\'\i sica do Porto e Departamento de F\'\i sica da
Faculdade de Ci\^encias da Universidade do
Porto, Rua do Campo Alegre 687, 4169-007, Porto, Portugal}
\author{E.P.S. Shellard}
\email[Electronic address: ]{E.P.S.Shellard@damtp.cam.ac.uk}
\affiliation{Department of Applied Mathematics and Theoretical Physics,
Centre for Mathematical Sciences,\\ University of Cambridge,
Wilberforce Road, Cambridge CB3 0WA, United Kingdom}

\begin{abstract}
We study the behaviour of cosmic string networks in contracting universes,
and discuss some of their possible consequences.
We note that there is a fundamental time asymmetry between defect network
evolution for an expanding universe and a contracting universe.
A string network with
negligible loop production and small-scale structure will asymptotically
behave during the collapse phase as a radiation fluid. In realistic networks
these two effects are important, making this solution only approximate.
We derive new scaling solutions describing this effect, and test them
against high-resolution numerical simulations.
A string network in a contracting universe, together 
with the gravitational radiation background it has generated, 
can significantly affect the dynamics of the universe both locally and 
globally. The network can be an important source of radiation, entropy 
and inhomogeneity.  We discuss the possible implications of these findings
for bouncing and cyclic cosmological models.
\end{abstract}
\preprint{DAMTP-2002-59}
\maketitle

\section{\label{intr}Introduction}

Cosmological scenarios involving oscillating or cyclic universes
have been know for a long time \cite{Tolman}, with interest in them varying
according to the latest theoretical prejudices or observational
constraints.  Recent interest has been associated with a cyclic extension
of the ekpyrotic scenario \cite{Steinhardt}.
A related result was
the realization \cite{Kanekar,Peter1,Peter2}
that the presence of a scalar field seems to
be necessary to make cosmological scenarios with a bounce
observationally realistic. And if scalar fields are present,
then one should contemplate the possibility of topological
defects being formed.

It is thought that the early universe underwent
a series of phase transitions, each one
spontaneously breaking some symmetry in
particle physics and giving rise to topological 
defects of some kind \cite{Kibble1,Book}, which in
many cases can persist throughout the subsequent evolution
of the universe.

In the present work we study cosmic string evolution in a
collapsing universe, following up on and generalizing the results
of \cite{Contracting},
and discuss in much greater detail some implications of the
presence of cosmic strings (and cosmic defects in general)
for bouncing universes. In a bouncing universe scenario the 
properties of the universe in the expanding phase depend on 
physics happening in a previous collapsing phase (before the 
bounce). For this reason, if defects do exist in these models,
it is crucial to understand their evolution and consequences
in both the expanding and collapsing phases.
Up to now all these studies, be they
analytic \cite{Kibble,Austin,Martins1,Thesis,Martins3}
or numerical \cite{Bennett,Allen1,Moore},
have only been undertaken for the expanding case, and it is
clear that while some results may be expected to carry over to
the contracting phase, some others clearly won't.
This will have consequences not only for the standard string seeded (or
hybrid) structure formation scenario \cite{Avelino2},
but also for other `non-standard' scenarios involving defects,
such as the production of adiabatic and nearly Gaussian
density fluctuations \cite{Adiabatic,Gaussian}, or
those involving anisotropic or inhomogeneous
universes \cite{Fossils,Inhomog}.

In particular, we expect that cosmic strings will become
ultra-relativistic, behaving approximately like a radiation
fluid. This means that a cosmic string network, both directly and through
the gravitational radiation emitted by the small loops it produces, will
soon become a 
significant source of entropy (and also of
inhomogeneity). Hence a cosmic string network
is a further problem for
cyclic universes if a suitable and efficient mechanism for diluting the
entropy is not available.

We should point out at the outset that if/when during the collapse phase
one reaches the Hagedorn temperature, one expects the string network to quickly
dissolve. However this is largely irrelevant for the points being made in
this paper: the radiation, entropy and anisotropy produced by the network
will obviously still be left behind if the network does dissipate. On the
other hand, it need not be the case that the Hagedorn temperature is reached
or, more specifically, that the collapse has to continue all the way to
the 'Big Crunch' (where there is no known sensible description
of the physics involved). Cosmological models do exist where the bounce takes place at finite size. Indeed, such models seem to be relatively common
in scenarios with extra dimensions and scalar fields, 
although this issue is somewhat debatable.

The outline of this paper is as follows. In Sect. \ref{cyc} we present
a very brief overview of previous work on cyclic universes,
and discuss further motivation for our work.
Then in Sect.\ref{stringev} we introduce the basic
dynamical properties of cosmic string networks, and after
a `warm-up' example of a circular string loop we successively
discuss analytic scaling solutions in various regimes
in a contracting universe. In Sect. \ref{numerics} we
describe high-resolution numerical simulations of string networks
in contracting universes using the Allen-Shellard string code \cite{Allen1},
and then compare and contrast our analytic and numerical results.
Possible cosmological consequences of these results are discussed
in Sect. \ref{conseq}, and finally we present some conclusions
in Sect. \ref{conc}. Throughout the paper we will use units
in which $c=\hbar=1$.

\section{\label{cyc}An overview of bouncing cosmological models}

Oscillating universes arise naturally
as classical exact solutions of the Einstein equations,
and have been explored in several contexts in
an attempt to solve some of the cosmological enigmas.
For example, they could conceivably solve the
flatness and horizon problems. For oscillating universes
whose size at maximum expansion increases in each cycle the flatness
problem may be solved because it is ever less likely
for an observer to find himself
in a non-flat region. Provided there is a causal
correlation for the microphysics at the bounce,
the age for cyclic universes
may also be large enough to solve the horizon problem.

However, oscillating models are not without their caveats.
The first of these was pointed out a long time ago
by Tolman \cite{Tolman}. By assuming a closed universe with
zero cosmological
constant in the context of the General Relativity theory,
he noted that entropy is generated at each cycle and
so the total entropy in the universe grows from cycle to cycle.
Hence the period of each cycle is larger that the previous one,
and extrapolating back in time one finds that the sum is finite.
In other words, the universe must still have had a beginning.
These points were also discussed by Rees \cite{Rees},
and formalized by Zel'dovich and Novikov \cite{Zeldovich}
and subsequently by others \cite{Markov,Penrose}.

A thorough analysis of oscillating universe solutions was
first carried out by Barrow and Dabrowski \cite{Barrow}.
They showed that in closed universes filled
with a perfect fluid of matter and/or
radiation, with total entropy increasing from cycle to cycle,
a positive cosmological constant or any non-zero stress violating 
the strong energy condition will eventually halt the oscillations
after a finite number of cycles.
In particular when the cosmological constant dominates the universe
approaches the de Sitter solution.
Dabrowski \cite{Dabrowski} subsequently generalized
the analysis to include negative pressure matter 
which violated the strong energy condition, and also provided
a somewhat simplistic description of the scalar field case. 

Looking for an alternative model to inflation
that could explain the scale invariance of the
observed power spectrum of fluctuations,
Durrer and Laukenmann \cite{Durrer} 
presented a model of a closed universe dominated
by radiation or with an intermediate matter-dominated
period where small black holes could have been formed.
They \textit{postulated} that density perturbations produced
gravitational entropy which would have to be transformed into radiation
entropy at the bounce. This could explain most of 
the present day radiation entropy without having overproduction,
provided one also assumed that perturbations never became
strongly non-linear in the previous cycle (as the mass fluctuation
would in fact generally diverge at the big crunch).

More recently, a detailed analysis by Peter and
Pinto-Neto \cite{Peter1} has shown quite generically
that no detectable bounce
is observationally allowed in any universe which around the
epoch of the bounce is described by Einstein
gravity together with hydrodynamical fluids, since under those
conditions scalar density perturbations would become
non-linear well before nucleosynthesis. However, this
`no-bounce conjecture' can in some circumstances be evaded by adding
a free scalar field which can dominate the universe
during the bounce \cite{Peter2}.

One of the earliest discussions of bouncing universes with scalar
fields was by Hawking in the context of 
the no-boundary proposal \cite{Hawking}; of course, many of the 
discussions noted earlier are derivative of Hawking's cosmological
singularity theorems.  Early in the development of quantum cosmology, he 
studied a simple chaotic inflation model and suggested that his boundary conditions
selected only a special set of time-symmetric and eternally bouncing trajectories.  
Subsequent work, however, showed that these special bouncing 
states were not generic \cite{Page, Laflamme}, with typical 
universes diverging from what are unstable trajectories  \cite{Cornish} 
and re-collapsing to singularities.

Kanekar \textit{et al.} \cite{Kanekar} have also considered a massive 
scalar field in a closed universe for the Einstein theory. This is a possible
alternative mechanism which can still lead to an increase in the volume
of an oscillating universe without requiring entropy production.
They suggest that the oscillations of the
universe force the scalar field to also
oscillate about the minimum of its potential.
The work done by or on the scalar field
creates an asymmetry between the expansion
and collapse epochs which can result in the
increasing volume of the universe. It was also argued that
the presence of other matter fields will not change this picture,
provided that the interactions between the matter and scalar fields
are sufficiently weak.

Finally, we should mention the cyclic
extension \cite{Steinhardt} of the ekpyrotic
model \cite{Steinhardtetal,Steinhardtetal1}.
Here the contraction and expansion phases correspond to
the epochs before and after the collision of two three-branes
along a fifth dimension. During the expansion there are successive
periods of radiation, matter and vacuum domination (the latter
of which can dilute the entropy, black holes and other debris
produced during each cycle), and the inter-brane potential can be carefully
chosen to ensure the cyclic behaviour. Hence the model
can potentially solve many of the cosmological 
issues raised above. However, it appears questionable whether
density perturbations in this model, matched across the bounce, can
 reproduce the Harrison-Zel'dovich
spectrum, as has been pointed out by
a number of authors \cite{Finelli,Lyth,Martin1,Martin2}.

In the present work, motivated by the
Kanekar \textit{et al.} \cite{Kanekar} results,
we replace the scalar field by a cosmic string network, 
which is also expected to display an asymmetric behaviour
between the contraction and expansion epochs.
In particular, while during expansion a cosmic
string network will quickly evolve towards a linear scaling
regime \cite{Martins3} (except in very particular circumstances),
we shall see that this is not the case during a phase of
collapse. Indeed, in this case, a string network
where loop production and small-scale `wiggles' were negligible
would asymptotically behave like a radiation fluid.
As it turns out, these two effects are relevant throughout the
collapse (in fact, even more so than in the expanding phase),
and this solution is only approximate.

On the other hand, there is further radiation being produced: since
cosmic string loops decay gravitationally, the enhanced loop production
will give rise to an enhanced gravitational radiation background.
All in all, a cosmic string network will add a significant contribution,
in the form of radiation, to the energy (and hence also entropy)
budget of a contracting universe, which will become ever more
important as the contraction proceeds. In the following sections we
will demonstrate this behaviour and discuss its cosmological
consequences.

\section{\label{stringev}Cosmic string evolution: basics and analytic methods}

In this section we discuss the basic dynamical properties of cosmic string
networks, using both analytic and numerical methods, and use these tools
to characterize their behaviour in bouncing universes. Both the
analytic \cite{Martins1,Martins2,Open,Condmat,Thesis,Avelino1,Wiggly,Martins3}
and the numerical \cite{Allen1,Avelino2,Moore}
tools which we shall use rely on previous work by some of the
present authors. In what follows we will limit ourselves to
describing the features that are directly relevant for our
analysis. We refer the reader to the original references for
a more detailed discussion and derivations of some key results.

\subsection{Basics of string dynamics}

In the limit where the curvature radius of a cosmic string
is much larger than its thickness, 
we can describe it as a one-dimensional
object so that its world history can be represented by a
two-dimensional surface in space-time 
(the string world-sheet)
\be\label{6}
x^\nu = x^\nu (\sigma^a)\,; \qquad
a = 0,1\,; \quad\nu = 0,1,2,3
\ee
obeying the usual Goto-Nambu action.
For simplicity we start by considering
the string dynamics in a flat FRW universe 
with line element
\be
ds^2=a^2(\eta)(d\eta^2 - {\bf d x}^2).
\ee
Identifying conformal and world-sheet times and imposing
that the string velocity be orthogonal
to the string direction (i.e. $\dot {\bf x} \cdot {\bf x}^\prime = 0$) 
the string equations of motion take the form
\bq
&&\ddot {\bf x}
+ 2{\cal H}\,\left(1-{\dot {\bf x}}^2\right)\,\dot {\bf x} =
\frac{1}{\epsilon} \,
\left( \frac{{\bf x}^\prime}{\epsilon}\right)^\prime
\label{strings}\\
&&\dot \epsilon + 2 {\cal H}\,{\dot {\bf x}}^2\,\,\epsilon = 0
\label{stringt}
\eq
where the `coordinate energy per unit length',
$\epsilon$, is defined by
\be\label{14}
\epsilon^2 = \frac{{\bf x}^{\prime\,2}}{1- {\dot {\bf x}}^2}\, ,
\ee
${\cal H}={\dot a}/a$, and dots and primes are derivatives with respect to 
the time and space coordinates.
Note that if the universe is not spatially flat
then the microscopic string equations of motion
(\ref{strings}-\ref{stringt}) will have additional curvature
corrections---see \cite{Thesis,Avelino1,Martins3}.

The evolution of the scale factor is described by the Friedmann equation 
\be
{\cal H}^2+K={\cal H}_0^2 (\Omega_m^0 a^{-1} + \Omega_r^0 a^{-2} + 
\Omega_\Lambda^0 a^2)
\label{frdeq}
\ee
where we have considered only three contributions to the energy density of the 
universe, namely matter, radiation and a cosmological constant.
Note that this assumes that the cosmic string density is comfortably
subdominant and need not be included in the Friedmann equation. Although
this is usually the case in most cosmological scenarios, it need not
happen all the time.

\subsection{A simple example: the circular loop}

We will now consider the evolution of a circular cosmic string loop in a 
cyclic universe. Although realistic loops chopped off by the network are
of course highly irregular (and, crucially, have a considerable
amount of small-scale `wiggles' \cite{Allen1,Wiggly,Moore}),
the circular solution is still a useful starting point and
can provide some intuition for things to come \cite{Martins2}.
In this example we will assume for simplicity that the universe is
spatially flat and that there is a negative cosmological constant which
makes the universe collapse.

The loop's trajectory can be described by
\be\label{17}
{\bf x} = r(\eta) \, \left(\sin\theta, \cos\theta,0\right).
\ee
with $0\leq \theta\leq 2 \pi$.
We define $R=|r| a \gamma$ as the `invariant' loop radius 
(which is proportional 
to the energy of the loop), and  
$\gamma=(1-v^2)^{-1/2}$, with $v=|{\dot r}|$
being the microscopic speed
of the loop. The microscopic string equations of motion
(\ref{strings}-\ref{stringt}) then become
\be
\ddot r=(1-{\dot r}^2) \left(-\frac{1}{r}-2H{\dot r}\right)\,,
\label{26}
\ee
where $H={\cal H}/a$,
with the velocity equation being obviously obtainable form this,
or alternatively in terms of the invariant loop radius
\be
\frac{dR}{dt}=(1-2v^2)HR\, .
\label{evlolR}
\ee
This latter equation coincides with the averaged evolution equation for
$R$, obtained in the context of the velocity-dependent
one-scale model \cite{Martins2,Thesis,Martins3},
while for the averaged velocity equation this model yields
\be
\frac{d{\bar v}}{dt}=(1-{\bar v}^2)\left(\frac{k(
{\bar v})}{R}-2H{\bar v}\right)\, ,
\label{loopv}
\ee
where $k({\bar v})$ is the momentum parameter which is thoroughly discussed
in \cite{Martins3}.

Simple analytic arguments show that a loop whose initial
radius is much smaller than the Hubble radius will oscillate 
freely with a constant invariant loop radius and an average velocity
${\bar v}=1/{\sqrt 2}$. (Note that we are assuming units in which $c=1$.) 
On the other hand, once the collapse
phase begins, we will eventually get to a stage in which the physical
loop radius becomes comparable to the Hubble radius $ar\sim H^{-1}$ 
and then gets above it.
In this regime the loop velocity is typically driven towards unity 
$v\rightarrow 1$ and it is straightforward to show that the 
invariant loop length grows as $R \propto a^{-1}$ and the Lorentz factor as
$\gamma\propto a^{-2}$.  Despite its growing energy $R$, the actual physical 
loop radius $ar = R/\gamma \propto a$,
so the loop shrinks with the scale factor
and inexorably follows the collapse into 
final big crunch singularity.   

Importantly, note that this relativistic final 
state for a loop in a collapsing universe is generic and quite 
different to the initial condition
usually assumed for super-horizon loops in the expanding phase (created, say,
at a phase transition).  In the latter case, 
the loops begin with a vanishing velocity
which only becomes significant when
each of them falls below the Hubble radius.
Such evolution cannot be reproduced in 
reverse during the collapsing phase without
fine-tuning the velocity as the loop 
crosses outside the Hubble radius.  This simple fact introduces a 
fundamental time asymmetry for string evolution in a cyclic universe.
In fact, something analogous happens for all other topological defects. 

\begin{figure}
\includegraphics[width=3.5in,keepaspectratio]{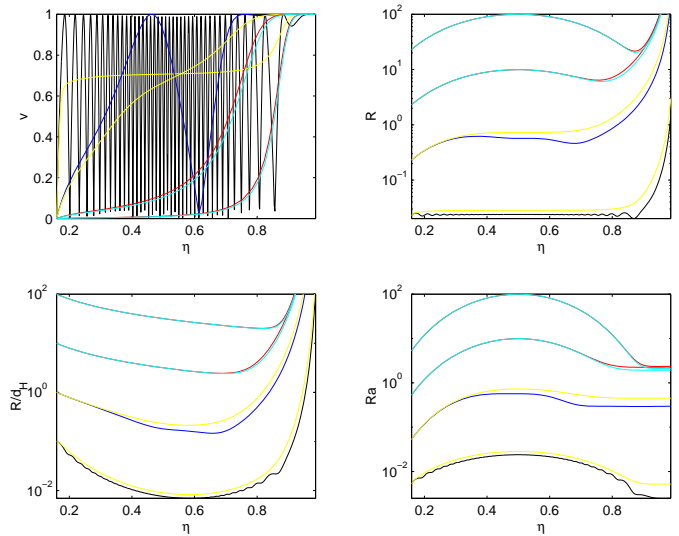}
\caption{\label{loopm}Comparing the microscopic (darker curves) and
averaged (lighter curves) evolution for four circular loops in a
matter-dominated, closed universe. The conformal time was chosen
so that the big bang occurs at $\eta=0$ and the big crunch at $\eta=1$.
Displayed are the loop velocity $v$ (top left) and invariant radius
$R$ (top right), as well as the ratio of the loop and Hubble radii
$R/d_H$ (bottom left) and the product of the loop radius and the
scale factor $Ra$ (bottom right),
showing the asymptotic $R\propto a^{-1}$ behaviour.}
\end{figure}

The analytic expectations for our circular loop solution 
have been confirmed by a numerical study, see Fig. \ref{loopm}. 
Here we plot both the microscopic and
averaged loop sizes and velocities, for four loops of different sizes
in a universe filled with matter (the radiation case being analogous,
except for a slight difference in the timescale required to reach
this asymptotic regime).
Obviously a loop which has a considerable size relative to the
horizon to begin with will not oscillate. We can see that the
averaged quantities provide a very good description of the dynamics.

The main caveat with this solution is that it can not account for
the presence of small-scale structures, whose build up we
in fact expect to be enhanced in the collapsing phase.
One expects that the presence of
small-scale wiggles can significantly delay the onset of this
asymptotic regime---an expectation that we will confirm in what follows.
              
\subsection{\label{analytics}String network evolution: analytic expectations}

Two different but complementary approaches are available
to study the evolution of a cosmic string network: one can resort to
large numerical simulations \cite{Bennett,Allen1,Moore} (which are
intrinsically difficult and time consuming, as one is dealing with highly
non-linear objects), or one can develop analytic
tools \cite{Kibble,Martins1,Martins2,Thesis,Wiggly,Martins3} which provide an
averaged (or `thermodynamical') description of the basic properties
of the network. In what follows we shall briefly describe the best
motivated of these analytic models, the velocity-dependent one-scale (VOS)
model \cite{Martins1,Martins2,Thesis,Martins3}, and try to use it to deduce
the basic properties of a cosmic string network during a phase
of contraction. In the following subsection we will test these
against numerical simulations.

\subsubsection{The VOS model}

The VOS model describes the string dynamics in terms of two `thermodynamical' 
parameters: the string RMS velocity, $v_\infty$, defined by 
\be
v_\infty^2\equiv<{{\bf{\dot x}}}^2>=\frac{\int{\bf{\dot x}}^2\epsilon 
d\sigma}{\int \epsilon d\sigma}
\ee
and a single length scale, $L$, which can be variously interpreted as
the long string correlation length, its curvature radius, or simply a measure
of the energy density (see below). It is important to note that in the
context of this model, all these three length scales are assumed to
be identical \cite{Thesis}. Of course this assumption is not always
realistic and models exist where it is
relaxed \cite{Austin,Thesis,Wiggly}, but it
must be assumed if one wants to
use the model in this formulation. The string network is thus assumed to be
a Brownian random walk on large enough scales, characterized by a correlation
length $L$. Hence one can simply relate it with the energy density in
long strings as
\be
\rho_\infty=\frac{\mu}{L^2}\,,
\label{29}
\ee
where $\mu=\mu_0$ is the string mass per unit length.  Note that given 
that we will be considering relativistic velocities, the commonly used
`correlation length' $L$ is really a measure of the invariant string
length or energy, rather than the typical  
curvature radius of the strings.  By including the appropriate 
Lorentz factor $\gamma_\infty = (1-v_\infty^2)^{-1/2}$ for the long
strings, we can denote the physical correlation length  
by 
\be
L_{\rm phys} = L\gamma_\infty^{1/2}\,.
\label{physcorrel}
\ee
Note also that in defining
(\ref{29}) we are taking the string mass per unit length 
to be a constant.  More generally, if one wanted to
explicitly account for the presence of small-scale `wiggles' one would
need to define a varying `renormalized' mass per unit
length \cite{Thesis,Wiggly}, which would then be related to the `bare' mass by
\be
\mu={\tilde \mu}\mu_0\, .
\label{wiggly}
\ee

\begin{figure}
\includegraphics[width=3.5in,keepaspectratio]{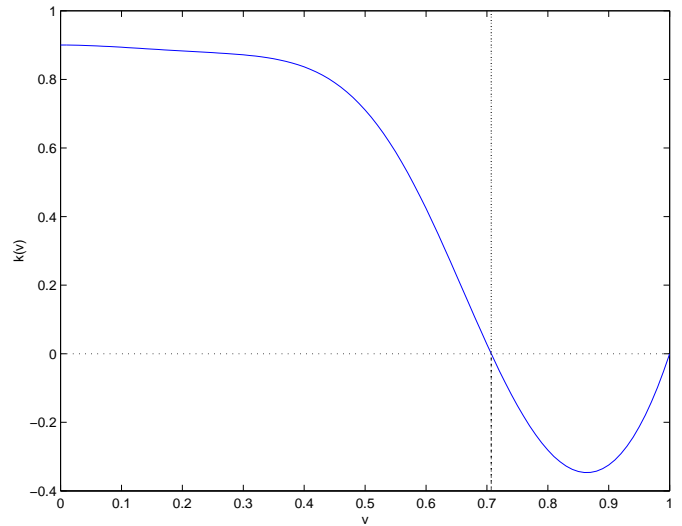}
\caption{\label{kappa}The momentum parameter $k(v)$ in the velocity-dependent
one-scale string evolution model. Notice that it vanishes at $v^2=1/2$
and $v^2=1$, and it is negative in between these two values.}
\end{figure}

With this assumption 
the VOS model has one phenomenological parameter ${\tilde c}$,
commonly called the loop chopping efficiency, which describes the
rate of energy transfer from the long-string network to loops.
The evolution equations then take the following form \cite{Martins1,Martins2}
\bq
&&2\frac{dL}{dt} = 2(1+ v_\infty^2)H L + {\tilde c} v_\infty
+8{\tilde \Gamma}G\mu v^6_\infty\label{50},\\
&&\frac{d v_\infty}{dt} = (1-v^2_\infty)\left(\frac{k(v_\infty)}{L}
-2 H v_\infty \right) \,.
\label{51}
\eq
The final term in the evolution equation for the correlation length
describes the effect of gravitational back-reaction.
Notice that we are not including in either equation additional
terms arising from friction due to particle scattering \cite{Martins3},
which could conceivably be important during the final stages of collapse.
We shall return to this point below.
Here $k(v_\infty)$ is the momentum parameter, which is thoroughly
discussed in \cite{Martins3}, and whose dependence on the string
velocity,
\be
k(v)\equiv \frac{2\sqrt{2}}{\pi}(1-v^2)(1+2\sqrt{2}v^3)
\frac{1-8v^6}{1+8v^6}\,
\label{kkvvv}
\ee
is shown in Fig. \ref{kappa}. Notice that this is
positive for $0<v^2<1/2$ and negative for $1/2<v^2<1$; this turns out to
be crucial for some of what follows.

As a final remark, we point out that these evolution equations are
approximate, and are neglecting higher-order terms \cite{Thesis}. It is
in principle possible that in this particular case ultra-relativistic
corrections may become significant, for example there could be powers of
$\gamma$ in various terms above. On the other hand, one
physically expects that the velocity increase will also be accompanied
by an enhanced build-up of small-scale `wiggles', for which one might need
further corrections \cite{Thesis,Wiggly}.
>From a `phenomenological' modelling point of view,
the latter are much harder to calculate than the former, and as far as one
knows this can only be done in the context of proper wiggly models of
string evolution \cite{Wiggly,School}. Thus in the present
work we chose not to further
complicate the model with these hypothetical corrections. We shall discuss
these issues further below.

\subsubsection{The ultra-relativistic regime}

The first point to notice about eqns. (\ref{50}-\ref{51}) is that
when the contraction phase starts and the Hubble parameter becomes
negative the velocity will tend to increase---the Hubble term becomes
an acceleration term, rather than a damping one. This will be compounded
be the fact that once one is beyond $v^2=1/2$ the momentum term in
the velocity equation also changes sign (\textit{cf.} Fig. \ref{kappa}),
but it is clear that given the constant increase in the magnitude of
$H$ this term will gradually prevail (more on this in the following
sub-section).

Thus one can expect that, just as in the simple case of the circular loop,
the string velocity will gradually tend towards unity. In this approximation,
and neglecting for the moment the loop production and gravitational
back-reaction terms, the evolution
equation for the correlation length easily yields
\be
L\propto a^2\,.
\label{basicl}
\ee
Note that
this is the same overall scaling law for the string network as that in a 
radiation-dominated expanding universe; the string network 
effectively behaves like a radiation component.  In terms of the physical correlation 
length (\ref{physcorrel}) we have $L_{\rm phys} \propto a$, which is as 
if the strings were being conformally contracted (except for their rapidly 
growing velocities).

However, there are several factors that must be considered which complicate
this simple state of affairs. First, there is the issue of
loop production. Under the above assumptions on velocity, but putting the
loop production term back in the correlation length equation, one finds
the following approximate solution in the radiation-dominated case,
\be
L_{rad}=\left(L_{max}-\frac{{\tilde c}}{2}\ln{a}\right)a^2\,
\label{asymprad}
\ee
and, in the matter era, 
\be
L_{mat}=\left[L_{max}+\frac{{\tilde c}}{2}\left(a^{-1/2}-1\right)\right]a^2\,
\label{asympmat}
\ee
where $L_{max}$ is the string correlation length at the
time of maximal size of the universe, and the scale factor at that time
was chosen to be unity (so the logarithmic
correction in the first case is positive).
Hence we see that, if ${\tilde c}$ remains constant (or is
slowly varying), asymptotically the scale factor dependent terms will
dominate, so that we expect $L\propto a^2 \ln{a}$ in the radiation era,
and $L\propto a^{3/2}$ in the matter era. Again we note that the latter
is also the scaling law for the correlation length in the
matter-dominated, expanding universe (but again the behaviour of the
velocities is very different in the two cases). This highlights the
different roles played by loop production in the scaling behaviour of
a cosmic string network in the radiation and matter eras, a point which
has been noticed long ago \cite{Kibble,Thesis} in the usual expanding
case.

A strong argument can be made, however, for an important relativistic correction
to the loop production term in
the evolution equation for the correlation length. In the simplest form of the 
VOS model there is a simple identification between the correlation 
length, $L$, and the physical distance $L_{\rm phys}$ which a string segment 
is expected to travel before encountering another segment of the same size 
forming a loop in the process. However, taking into account the Lorentz factor
in (\ref{physcorrel}), 
this means that the probability $dP$ that a string segment 
will encounter another segment in a time  interval $dt$ should be  
given approximately by
\be
dP = - \frac{d \rho_\infty}{\rho_\infty}=2 
\frac{d L}{L} \sim \frac{v_\infty dt}{L_{ph}} \sim
 \frac{v_\infty dt}{\gamma_\infty^{1/2} L}\,.
\label{probloop}
\ee
One would expect ${\tilde c} \propto \gamma_\infty^{-1/2}\propto a$ thus 
driving ${\tilde c}$ 
rapidly towards zero and asymptotically yielding our simple solution 
$L \propto a^2$ both in the 
radiation and matter eras. 

Of course, during re-collapse we expect that 
${\tilde c}$ will depend on a number of other properties of the 
string network such as the enhanced build-up of small scale structure due to
the contraction.  Eventually, however, the Hubble radius will fall below 
even the length scale of wiggles on the string after which our asymptotic 
solution $L\propto a^2$ should be valid.  In what follows, we shall consider
the two well-motivated cases,  first  ${\tilde c}=\hbox{const.}\ne 0$,
and secondly ${\tilde c}=0$, the probable asymptotic case.

\subsubsection{Approaching the ultra-relativistic regime}

Returning to our analytic solutions for the constant loop
production case (\ref{asymprad}) and (\ref{asympmat}), we can use 
the velocity equation to find an
approximate solution for the evolution of the velocity as it
approaches unity. Taking the
simplifying assumption that the momentum parameter is slowly varying,
one can arrive at the following implicit solution
\be
1-v^2\propto a^4\exp{\left[\frac{2k(v)}{\lambda}\int
\frac{a^{1/\lambda}da}{L(a)}\right]}\,,
\label{asympv}
\ee
where $\lambda=1$ in the radiation era and $\lambda=2$ in the matter era.
Substituting (\ref{asymprad}-\ref{asympmat}) one respectively obtains
\be
1-v^2_{rad}\propto a^4\left(-\ln a\right)^{4k(v)/{\tilde c}}\\,
\label{asympvvrad}
\ee
in the radiation-dominated case, and
\be
1-v^2_{mat}\propto a^{4+2 k(v)/{\tilde c}}\,
\label{asympvvmat}
\ee
in the matter-dominated case. Recall that the momentum parameter $k(v)$
is negative in the ultra-relativistic regime. Hence in the limit where
$v\to 1$ and therefore $k\to 0$ the asymptotic solution would have the
form
\be
\gamma^{-2}\propto (1-v^2)\propto a^4\,.
\label{asympvvv}
\ee
The presence of the momentum corrections, which phenomenologically
account for the existence of small scale structures on the strings,
imply that convergence will be slower than this power.

\subsubsection{Friction from particle scattering}

So far we have been neglecting the contribution of friction due to particle
scattering for the evolution of the cosmic string network. The reason is that
previous work \cite{Martins1,Martins2,Thesis,Condmat} has shown that for
heavy strings (say, those formed around the GUT scale) this will only be
significant very close to the Hagedorn temperature, that is, for temperatures
close to those of the string-forming phase transition.
For the rest of
the cosmic history of the network, this friction is always subdominant
relative to the damping generated by the expansion itself. However, for
light strings (say, those formed around the electroweak scale) friction is
significant for a longer period, and hence it is important to discuss how
our previous results might be changed in this case.

One can show \cite{vilfric,Book} that the effect of friction due to particle
scattering may be characterized by a friction length scale
\be
\ell_{\rm f}=\frac{\mu}{\beta T^3}\approx \frac{\eta ^2a^3}{\beta T_0^3}\,,
\ee
where $T$ is the temperature ($T_0\sim 1$meV today), $\eta$ is the symmetry-breaking 
scale of the string and $\beta$ is a phenomenological parameter counting
the number of particle species interacting with the strings ($\beta \sim {\cal O}(1)$).
Including this term in the string equations of motion (\ref{50}-\ref{51}) is
then fairly straightforward---see \cite{Martins2,Thesis} for a derivation.
One finds
\bq
&&2\frac{dL}{dt} = 2(1+ v_\infty^2)H L + \frac{L}{\ell_{\rm f}}v_\infty^2+ 
{\tilde c} v_\infty +8{\tilde \Gamma}G\mu v^6_\infty\label{50elf},\\
&&\frac{d v_\infty}{dt} = (1-v^2_\infty)\left[\frac{k(v_\infty)}{L}
-\left(2 H +\frac{1}{\ell_{\rm f}}\right)v_\infty \right] \,.
\label{51elf}
\eq
The epoch at which the frictional force becomes subdominant is given by
\be
\frac{t_*}{t_{hag}}=\left(\frac{a}{a_{hag}}\right)^2\approx \beta^2\left(\frac{45}{16\pi^3{\cal N}}\right)^{1/2}\frac{1}{G\mu}\,, 
\label{tstarexp}
\ee
where ${\cal N}$ counts the total number of effectively massless
degrees of freedom in the model
(hence ${\cal N}=106.75$ for a minimal GUT model, but it can be more than an
order of magnitude higher for particular extensions of it).

\begin{figure}
\includegraphics[width=3.5in,keepaspectratio]{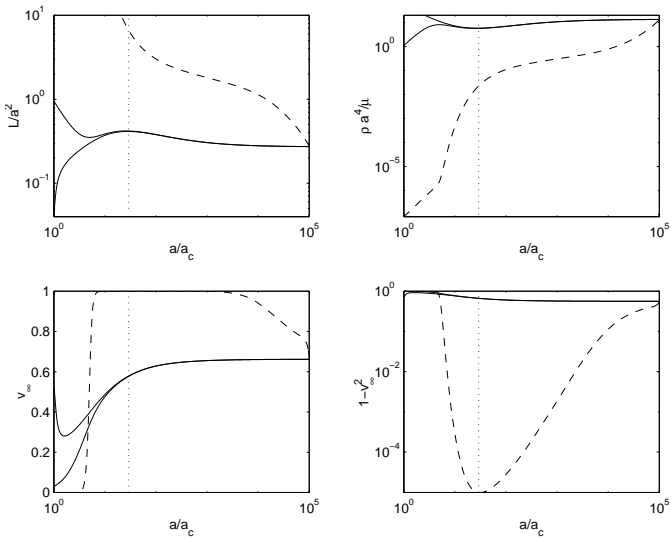}
\caption{\label{figfric}Illustrating the effect of friction due to particle
scattering in the expanding and contracting phases. A GUT-scale string network
is evolved forward for 10 orders of magnitude in time starting at the epoch
of formation (solid lines). The two lines correspond two different initial
conditions, characteristic of first and second order phase transitions.
At this point the evolution is switched to the contracting phase, and the
network evolved back until the Hagedorn temperature.
(dashed lines). The dotted line marks the end/beginning of the
friction-dominated epoch, $a_*$ (see the discussion in the text).}
\end{figure}

One can then proceed as before and look for scaling solutions for the particular
case when this extra frictional term dominates over the damping due to the Hubble flow.
Again we start by considering the simplest case where both loop production and
gravitational back-reaction are neglected. Assuming a relativistic network $v\approx 1$, 
the criterion for friction dominating the velocity evolution is 
$\ell_{\rm f}^{-1} > 2H$ which 
in the collapsing radiation era occurs at when the scale factor
again reaches \ref{tstarexp}, or alternatively in terms of the
time $t_*$ \textit{before} the big crunch
\be
t_{\rm c} - t_* \approx (G\mu)^{-2} m_{\rm pl}^{-1}\,, 
\label{tstar}
\ee
or, still equivalently, the temperature $T_* \approx \eta^2/m_{\rm pl}$
with $a_* \approx \eta^2/(m_{\rm pl} T_0)$. This expectation
can be confirm by solving the above evolution equations numerically,
as shown in Fig. \ref{figfric}. For $t>t_*$ and 
while the string remains relativistic ($v\approx 1$), 
it is easy to see that exponential brakes are 
applied to the momentum with 
\be
\gamma v = \gamma_* v_* \exp{ \frac{1}{2}\left[   1- \frac{a_*}{a}\right]}\,.
\label{expbrakes}
\ee
Even if $\gamma_*v_*$ was extremely large at $t=t_*$, the strings would soon 
be driven non-relativistic
when the scale factor had shrunk by only the logarithmic factor 
$a_{\rm nr}/a_* \approx 1/\ln(\gamma_*v_*)$.

From this point onwards, the velocity evolution becomes analogous to friction domination
in flat space, with Hubble anti-damping irrelevant and the small velocities 
simply reflecting a balance between the curvature scale and the 
friction length:
\be
v \approx \frac{\ell_{\rm f}}{L}\,.
\label{fricvel}
\ee
When the velocity is driven away from unity $v<<1$ at $t_{\rm nr}$, the previous
scaling law $L \propto a^2$ breaks down.
Substituting (\ref{fricvel}) into the evolution equation for the correlation 
length, then yields the simple
asymptotic solution for $a << a_{\rm nr}$,
\be
L \propto a\,, \qquad\qquad v \propto a^2\,.
\label{friction}
\ee

The string network ends in friction domination by coming to a standstill in 
comoving coordinates and then being genuinely conformally contracted.
The energy density in the string network now behaves as $\rho_\infty \propto a^{-2}$,
so if it has not dominated the energy density before $t_{\rm nr}$, then it
never will before the next bounce or final big crunch.  This final demise for the 
network is expected.  As the background temperature and density approaches 
those 
of the original string-forming  
phase transition the strings should effectively dissolve back into the 
high density radiation background from whence they had come.

\subsection{Further improvements}

\subsubsection{Dynamical friction}

We now discuss some further contributions to the dynamics of the cosmic string
network in some detail, as well as a few caveats to this approach.
An effect related to friction from particle scattering  is
dynamical friction, first discussed in \cite{Garfinkle,Dynamical}.
Cosmic string dynamics will be damped by dynamical friction,   
the magnitude of the effect depending both on the nature of the 
background fluid, the string mass per unit length and the amount 
of small scale structure of the cosmic string network. 

For the simple case of a wiggly string oriented along the $z$ axis moving 
in a matter background this effect can be easily computed from the fact 
that in the rest frame of the string the velocity change in the direction 
perpendicular to the string experimented by a matter particle is given 
by \cite{Book}:  
\be
\delta v_y = -\frac{2 \pi G ({\tilde \mu} - {\tilde T})}{v_s \gamma^2}-
4 \pi G {\tilde \mu} v_s,
\ee
where $v_s$ is the string velocity with respect to the matter particles and ${\tilde \mu}$ and ${\tilde T}$ are the effective string mass and 
tension per unit length, respectively.
It should be obvious that the total momentum transferred to the particles in 
the $y$ direction vanishes because there is a cancellation due to particles 
passing above and below the string. This is not the case in the $x$ direction 
where we may calculate $\delta v_x$ from the fact that the value of the 
velocity of the mass particle should be the same before and after the interaction has taken place. This means that to second order in $\delta v_y$ we 
have $\delta v_x = -\frac{1}{2}(\delta v_y)^2/v_s$ so 
that
\be
\delta v_x = -\frac{2 \pi^2 G^2({\tilde \mu} - {\tilde T})^2}{v_s^3 \gamma^4}
 -8 \pi^2 G^2{\tilde \mu}^2 v_s-\frac{8 \pi^2 G^2({\tilde \mu} - {\tilde T}){\tilde \mu}}{v_s \gamma^2}\,.
\ee

The total momentum transferred to a particle in the $x$ direction in the rest 
frame of the string is then $m \gamma \delta v_x$. This means that 
the total momentum transferred to the string in the interval of time $dt$ 
in this frame is given approximately by 
\be
-\delta v_x \rho v_s d t \int_{-\xi}^{\xi} dz \int_{-\xi}^{\xi} dy \sim - \frac{3 v_s \delta v_x dt}{32 \pi G}\,,
\ee
where $\xi \sim H^{-1}$ is the string length and we have assumed the universe 
to be flat with background density $\rho=3H^2/(8 \pi G)$.
Making a Lorentz transformation to the rest frame of the fluid we finally 
obtain that the fraction, f, of the string momentum, $p={\tilde \mu} \gamma \xi v_s$, 
lost 
in one Hubble time 
is given by
\be
f \equiv - \frac{dp}{dt} \frac{H^{-1}}{p} \sim \frac{3 \pi G}{4}  \left( \frac{({\tilde \mu} - {\tilde T})^2}{4 {\tilde \mu} v_s^3 \gamma^4}
 +{\tilde \mu} v_s+\frac{({\tilde \mu} - {\tilde T})}{v_s \gamma^2}\right)\,.
\ee
We note that this simple analysis is linear on $\delta v_y$ which 
effectively means that we are assuming $G {\tilde \mu} \ll 1$ and  
$v_s \gg G ({\tilde \mu}-{\tilde T})$.

In the context of a collapsing universe the ultra-relativistic 
($\gamma \gg 1$) regime is the most relevant. In this regime we have 
$f \sim O(G {\tilde \mu} )$ during the matter dominated era. 
The effect of dynamical friction in a radiation background 
is less dependent on the small scale structure on the string and is 
simply given by\cite{Garfinkle,Dynamical} :
\be
f \equiv - \frac{dp}{dt} \frac{H^{-1}}{p} \sim  \frac{3 G {\tilde \mu}\gamma(1+v_s^2/3)}{32 \pi}.
\ee
In the ultra-relativistic regime we have $f \sim O(G {\tilde \mu} \gamma)$ 
during the radiation dominated era.  We note that these results correct 
equation typos (by a factor of $\gamma$) in refs. 
\cite{Dynamical, Contracting}. 

Hence, we conclude that in a collapsing universe and during the matter era 
dynamical friction is never able to slow the strings down significatively so 
long as $G {\tilde \mu} \ll 1$.  On the other hand, during the radiation era 
dynamical friction will have a bigger impact on string dynamics and could 
in principle halt the evolution of $\gamma$ at $\gamma \sim 
(G {\tilde \mu})^{-1}$. Having said this, it is important to keep in mind that 
this analysis assumes a homogeneous and isotropic background and so may not 
strictly apply in our case. Moreover, the fact that a significant amount of the 
momentum will be transferred from the strings to the background (specially 
in the radiation era regime with $\gamma \sim (G {\tilde \mu})^{-1}$) will in 
itself add to the anisotropies which naturally occur in our model.

\subsubsection{A string-dominated collapsing universe}

If the energy density in the cosmic string network becomes a
non-negligible contribution to the overall energy density of the
universe before $t_{\rm nr}$, then the Friedmann eqn. (\ref{frdeq}) must be modified to
include the string density contribution on the right-hand side.
Furthermore, it may even happen that the string network becomes
the dominant contribution to the dynamics of the universe, and
in that case, the scaling laws can be significantly
modified---this type of scenario was first
studied, for the case of expanding universes, in \cite{DomV,DomK}.

In our case, one finds that to zeroth order the radiation-like behaviour
of the network, given by Eqns. (\ref{basicl},\ref{asympvvv}),
is maintained. However,
if one calculates (using the same method as sketched above)
the first order corrections to this behaviour, one finds that they are
slightly different, namely
\be
L=\left(L_{max}-{\tilde c}\ln{a}\right) a^2
\ee
and
\be
1-v^2\propto a^{4+2k(v)/\Delta}\,,
\ee
where we have defined $\Delta^2=8\pi G\mu/3$, that is a measure of the
string energy scale.

Notice that even though the loop chopping efficiency ${\tilde c}$ still affects
the behaviour of the correlation length, it no longer affects the behaviour
of the velocities (at least
in an explicit way). In this case the corrections to
the $\gamma\propto a^{-2}$ scaling depend only on the momentum parameter
$k(v)$ and on $\Delta$.

Also notice that $\Delta$ is a very small number, \textit{e.g.}
$\Delta_{GUT}\sim3\times10^{-3}$ for a GUT-scale network and
$\Delta_{EW}\sim3\times10^{-17}$ for an electroweak-scale network.
The correction term to the Lorentz factor scaling law is still
becoming less and less
important as the velocity increases, since the momentum parameter
is approaching zero, but the fact that $\Delta$ is so small implies
that the convergence towards the asymptotic solution is slower in
this case---the more so the lighter the strings are. 

\subsubsection{Other effects and caveats}

In the analysis in section \ref{analytics} we have not explicitly included
the effect of gravitational back-reaction.  However,
 these solutions will still hold when
this is incorporated. Indeed, in the context of the VOS model
one can rigorously show \cite{Martins3} that although this
term will clearly affect the quantitative values of the parameters
in a given scaling solution, as well as the timescale needed for such
solution to be reached within a given accuracy, it can not affect
the existence of such solutions.

One relevant issue is that of loop reclassification. In the  
expanding phase a small loop chopped off by the string 
network will slowly decay into gravitational radiation. 
However, in the collapsing phase it is possible that an initially 
small loop becomes a large loop (that is with a size in the center 
of mass (CM) frame comparable to $H^{-1}$) before it can lose all 
its energy. The size of the loop in the CM frame is given 
approximately by:
$$
L(t)=L_0-\Gamma G \mu (t-t_0)
$$
where $t_0$ is the physical time when the loop was produced, 
$L_0$ is the initial loop size, and the numerical coefficient $\Gamma$ 
is independently of the loop size , but does depend on its shape. 
Numerical simulations have shown that $\langle \Gamma \rangle \sim 65$.
The lifetime of the loop is then:
$$
\tau \sim \frac{L}{\Gamma G \mu}.
$$
Hence, we see that for $\tau > H^{-1}$ the loop will not have enough
time 
to lose all its energy before the big crunch which means that loops
produced 
with sizes greater than $\Gamma G \mu H^{-1}$ will eventually re-join
the 
long string network. This effectively contributes to a smaller value 
of $\tilde c$.

A final important element not included in the above solution is the
presence of small-scale `wiggles' on the string network and the
loops \cite{Bennett,Allen1}. The momentum parameter $k(v)$ can to
some extent account for this, but only in a very simple and
phenomenological way. As discussed elsewhere in this article,
in the case of a contracting universe, we expect
that the effect of the small-scale structures on the string dynamics
should be proportionally larger than that in the case of an expanding
universe. Although one could attempt to use more elaborate
analytic models to model this \cite{Austin,Thesis,Wiggly,School},
we shall leave this for further work, and for the moment restrict
ourselves to the simple and more intuitive VOS model.

Having said that, and even
considering that the model has been extensively tested and shown
to be accurate, one should keep in mind that in this case it
is not expected to do as well as in the previously studied cases.
Apart from the issue of small-scale structures, it is also worth
emphasizing that the VOS model assumes that the long string network
has a Brownian distribution on large enough scales, which may not
be a realistic approximation in a closed, collapsing universe.
This point clearly deserves further investigation.
Indeed, given the non-trivial (fractal) properties of a defect
network even in the simplest linear scaling regime, it is quite
interesting to ascertain what are the statistical properties of the
network in a closed universe around the epoch of maximal volume.
We leave this topic for future analysis.

\subsection{The equation of state}

An approximate equation of state for a cosmic string 
network in the relativistic limit is easy to obtain (see for 
example Kolb and Turner \cite{KTurner}): 
\begin{equation}
p_s = (2v_s^2 - 1)\frac{\rho_s}{3}\,.
\end{equation}
This result can be obtained by taking an average over all possible
directions of the energy-momentum tensor of a straight string.
The generalization of this result for a domain wall network is also
straightforward \cite{KTurner}:
\begin{equation}
p_w = (v_w^2 - \frac{2}{3})\rho_w\,.
\end{equation}
The same asymptotic limit is obtained when $v \to 1$ for both cosmic
strings and domain walls: they behave essentially as radiation.
The caveat here is that this derivation explicitly assumes a `perfect gas' of
strings or walls: in other words, it assumes that
there are no dissipative effects, either due to the defect motion
or to defect interactions.
Another minor caveat is that one should have the gas in a
box much larger than the network correlation length,
which will not be the case in a closed universe around maximum
expansion, just like the Brownian assumption.

This result could also be obtained from energy-momentum considerations taking
into account that the comoving momentum should be proportional to
$a^{-1}$. For a point mass this means that $\gamma_p v_p \propto a^{-1}$,
for a straight string $a \gamma_s v_s \propto a^{-1}$ and for a planar
wall $a^2 \gamma_w v_w \propto a^{-1}$. This means that when $v \to 1$ we
should have respectively
\be
\gamma_p \propto a^{-1}
\ee
\be
\gamma_s \propto a^{-2}
\ee
\be
\gamma_w \propto a^{-3}\,.
\ee
Taking into account the variation of the
comoving volume in obtaining the density of each one of these objects
during the collapsing phase it is straightforward to show that
$\rho_{p,s,w} \propto a^{-4}$ which is just radiation-like behaviour.

\section{\label{numerics}String network evolution: numerical simulations}

With the caveats discussed above in mind, we are now ready to
study the numerical evolution of a cosmic string network.

\subsection{The numerical code and basic checks}

We have performed a number of very high resolution Goto-Nambu
simulations on the COSMOS supercomputer, using a
modified version of the Allen-Shellard string code \cite{Allen1}.

\begin{figure}
\includegraphics[width=3.5in,keepaspectratio]{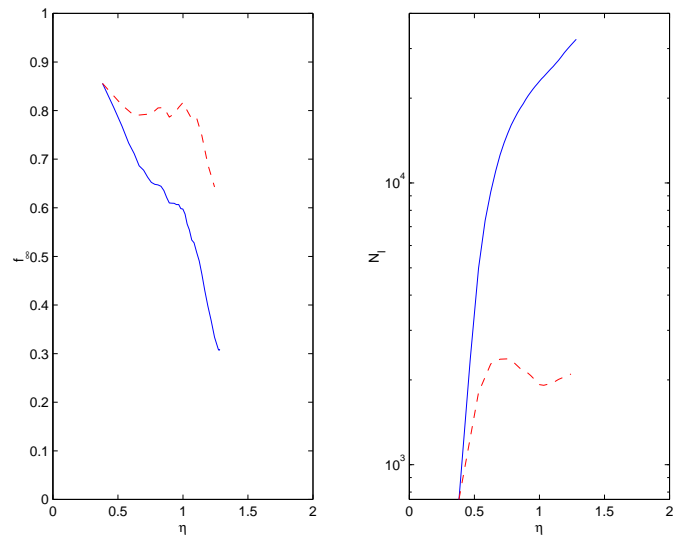}
\caption{\label{points}The same string network box is run without
(solid) and with (dashed) physical resolution point joining. This
crucially affects the loop production. The two panels show the fraction of the
total string energy that is in the form of long strings (left), and
the total number of loops in the simulation box (right), as a function
of the conformal time (defined such that maximal expansion occurs
at $\eta=1$ and the big crunch at $\eta=2$). This clearly shows
that high resolution is crucial in this case.}
\end{figure}

We have simulated a simplified scenario, where a radiation or matter
dominated universe is evolved in the expansion phase for a while,
to allow the initial conditions of the string box to be erased away
(for example, the initial string network is set up in a cubic lattice),
and then reversed the sign of (the square root of) the Friedmann
equation, thus forcing the universe to collapse. This setting
is numerically simpler to simulate than a closed universe, or
one whose collapse is induced by a negative cosmological constant,
but is still good enough to allow us to test the validity of the
above solutions.

In addition to the main set of simulations, we have performed a range
of other control simulations to test various numerical issues and
satisfy ourselves that our results had good enough numerical accuracy.
Part of these control simulations will be briefly described below.

We emphasize that given the expected enhancement of small-scale structure
on the strings, having very high resolution is a crucial factor,
especially in what concerns the loop population. In particular,
time-saving schemes like point joining are not accurate enough, even
if one keeps a constant physical resolution. This is illustrated in
Fig. \ref{points}, which shows the outcome of evolving the same
simulation box with and without a point joining approximation. One sees
that the number of loops produced is very different, and an even
more dramatic discrepancy can be noticed by looking at images of
the long strings in the box as the simulation evolves. We believe
that the main simulations shown below have a sufficient resolution
(in terms of points per correlation length) and dynamic range to
provide statistically significant results.

\begin{figure}
\includegraphics[width=3.5in,keepaspectratio]{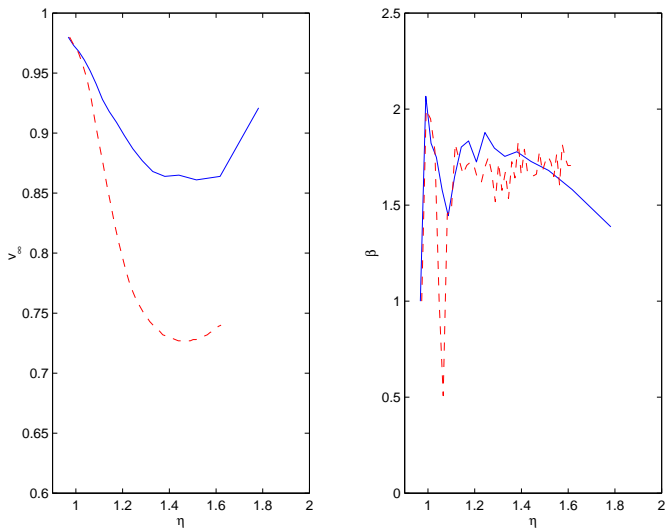}
\caption{\label{highv}Evolution of the long-string velocity (left panel) and
scaling law of $L\propto a^\beta$ (right panel) for simulations starting with
abnormally high velocity at the epoch of maximal size of the
universe. Both are shown as a function of conformal time (defined such that
maximal expansion occurs at $\eta=1$ and the big crunch at $\eta=2$).
Solid and dashed lines correspond respectively
to matter and radiation universes.}
\end{figure}

The first point to be established is that there is a solution with
the rough properties of the one described above, namely with the
velocity evolving from approximately $v\sim 1/\sqrt{2}$ (corresponding to
the usual scaling solution in the expanding case) to $v\sim1$
as the universe collapses, and that such solution is stable.
With this aim we have, among other tests,
carried out some simulations which start collapsing
after an expansion of a single time step, but which start out with
$v\sim 1$. The outcome of these is illustrated in Fig. \ref{highv}.

There is a very clear decrease of the velocity at early times, and an
apparent tendency for an increase at late times, though it must be
said that the numerical accuracy slightly deteriorates
at the very end of the simulation. Notice that the decrease is much stronger
for the radiation case. At the
same time, the power law dependence of the correlation length on the scale
factor, $L\propto a^\beta$, is approximately constant and in the
range $3/2<\beta <2$, except for very early in the simulation where
the network hasn't yet erased its initial conditions. Hence we can
conclude that these results are consistent with the existence of an
attractor solution of the type described above. Further evidence for
this convergence has been observed by other means, such as
starting simulations at maximal expansion with much lower velocity (say zero).

\subsection{Detailed scaling solutions}

\begin{figure}
\includegraphics[width=3.5in,keepaspectratio]{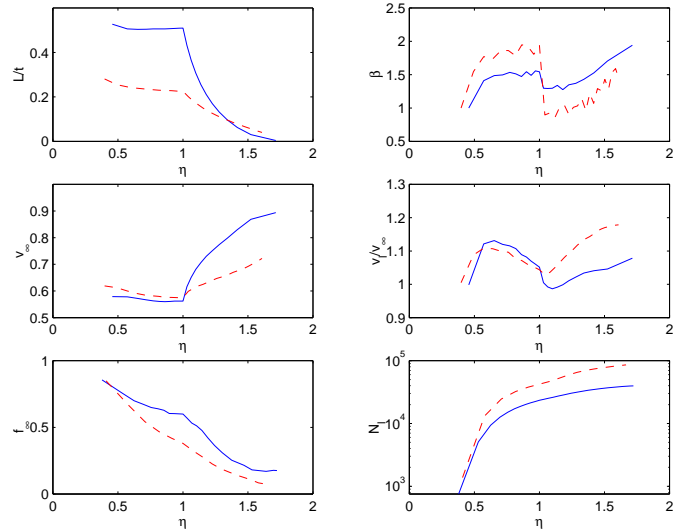}
\caption{\label{scalings}Basic properties of cosmic string networks
in the expansion and contraction phases, as a function of conformal time
(defined such that maximal expansion occurs at $\eta=1$ and the big crunch
at $\eta=2$). The solid represents a simulation in the
matter era, while the dashed line is for a radiation era run.
Plots successively show the behaviour
of the re-scaled correlation length $L/t$ (top left), the scaling law of
the correlation length relative to the scale factor
$L\propto a^\beta$ (top right), the velocity of the long string network
(middle left), the ratio of the loop and long string velocities (middle
right), the fraction of the string energy in the box in the form of
long strings (bottom left) and the total number of loops (bottom
right).}
\end{figure}

\begin{figure}
\includegraphics[width=3.5in,keepaspectratio]{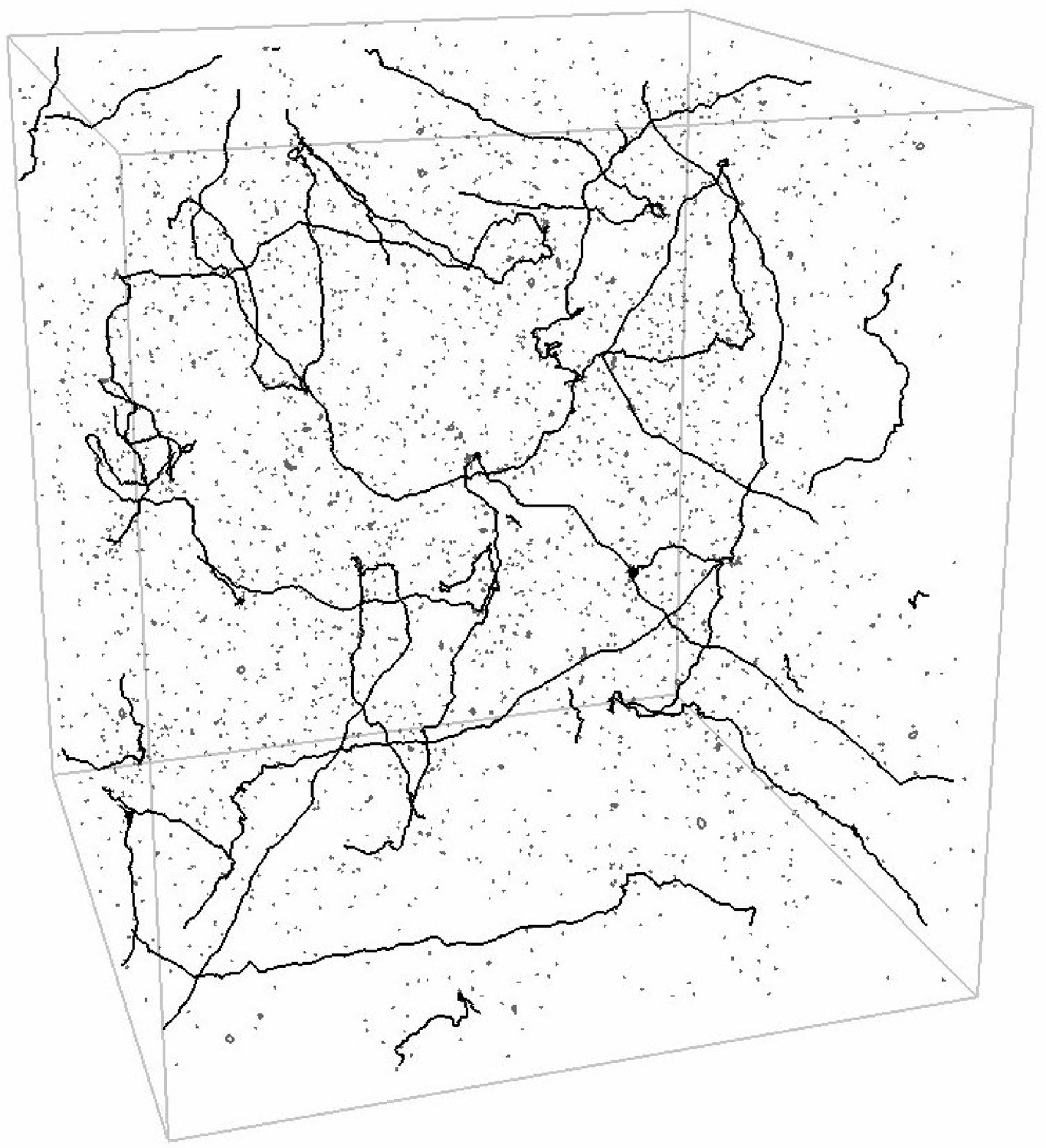}
\includegraphics[width=3.5in,keepaspectratio]{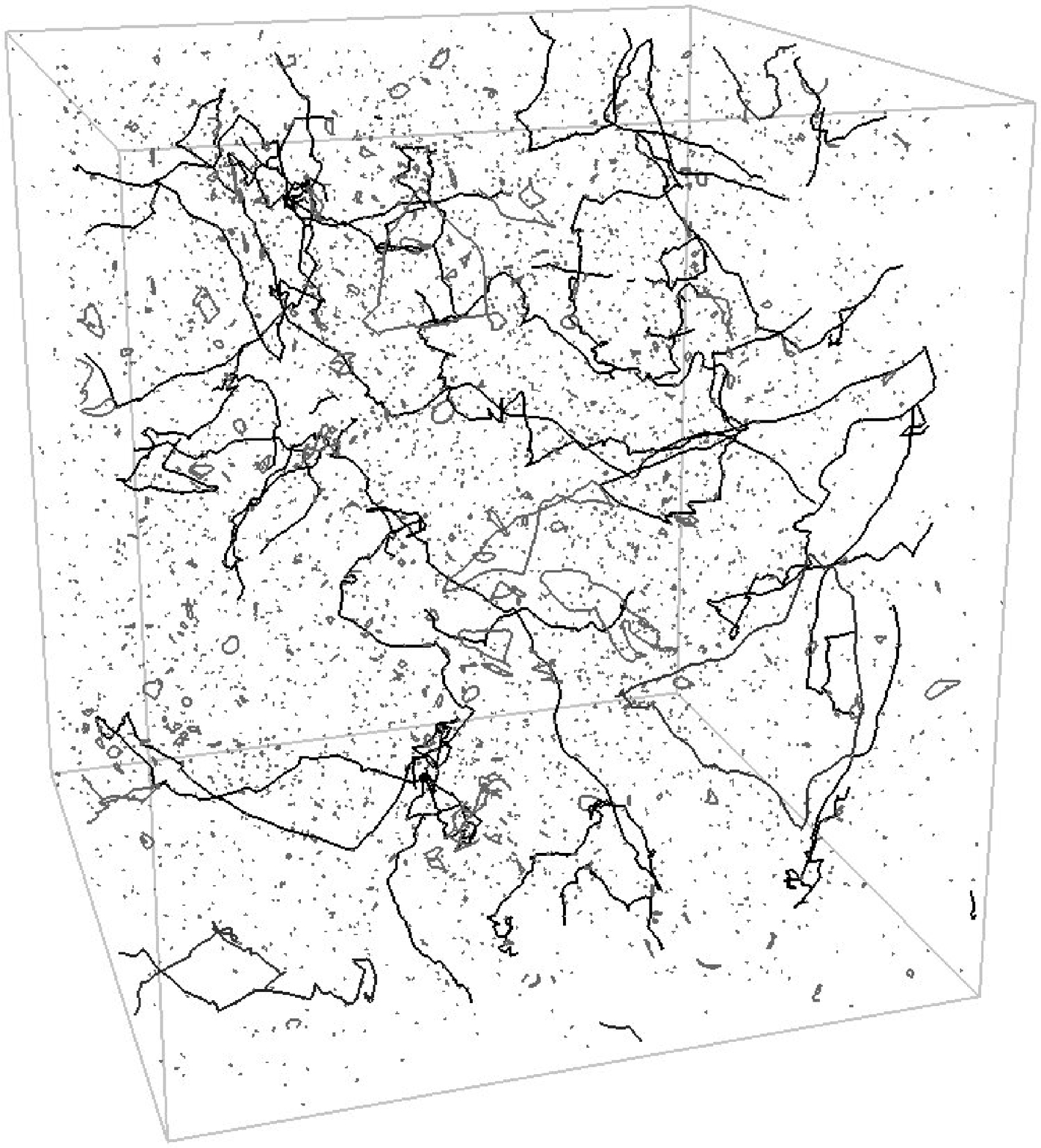}
\caption{\label{boxes}Two snapshots of the evolution of one of our
simulation boxes---they are typical representatives of string networks
in the expansion (top box) and collapsing (bottom box) phases, respectively.
Note that these snapshots are only a small fraction of the total
simulation box. In the first snapshot the horizon is about twice as
large as the box shown: in the second they are about the same size.}
\end{figure}

With some confidence in the robustness of our methods, we can now
proceed to study very high resolution simulations of the behaviour
of a string network in a universe with a period of expansion followed
by contraction. The result of two such simulations, for universes
filled with radiation and matter, is shown in Fig. \ref{scalings},
and two snapshots (typical of the expansion and contraction phases)
of a fraction of the simulation box in one
of the runs is shown in Fig. \ref{boxes}.

Notice that for times up to $\eta\sim0.5$ the network hasn't
yet erased its initial (lattice) conditions, so this period should
be disregarded when considering the scaling analysis.
We should also mention that for simplicity we are only
plotting every twentieth time step in the simulation, and also we are
not plotting the error bars associated with each data point. Depending on
the quantity, these are of the order of $10\% $, and are typically
larger at later than at earlier times.

During the expanding phase we confirm the usual linear scaling regime, namely
\be
L_{exp}\propto t\propto a^2,\qquad v_\infty=const.
\label{exprad}
\ee
in the radiation-dominated case, and
\be
L_{exp}\propto t\propto a^{3/2},\qquad v_\infty=const.
\label{expmat}
\ee
in the matter era.

As soon as the contraction phase starts, these laws are modified. As
expected, the velocity starts increasing, and the scaling of the
correlation length with the scale factor also drops, being approximately
constant to begin with, and then rising slowly. One can roughly
identify a transient scaling phase, valid in the period
$\eta\sim1.0-1.4$, where one approximately has
\be
L_{trans}\propto a
\label{expradtrans}
\ee
in the radiation-dominated case, and
\be
L_{trans}\propto a^{5/4}
\label{expmattrans}
\ee
in the matter era.
(These can not be easily recovered by analytic methods using the evolution
equations for the VOS model discussed above.)
Unfortunately, the extremely
demanding requirements in terms of resolution of the simulation
do not currently allow us to run simulations with longer dynamic
range to establish beyond reasonable doubt
whether this scaling law approaches
$\beta=2$, as predicted above.  However, there are strong indications 
that the networks are evolving towards this asymptotic regime, as 
shown by the relatively rapid climb of the exponent in Fig.~\ref{scalings}.

It is clearly noticeable that the velocity rises much faster in
the matter era than in the radiation era. It is also interesting to
point out that during the collapse phase the loop and long string
velocities are noticeably different, and this difference (which is more
significant in the radiation than in the matter case) increases with time.
The plot also shows an apparent difference in this velocity ratio
in the expanding phase, but this
is not significant. The initial lattice conditions of our simulations
do tend to give equal velocities to `long' strings and small loops, but
as they start evolving, the fact that the strings are on a lattice means
that there can be no inter-commutings for the first few time steps, and this
artificially makes the long string velocities fall behind those of the loops.
Eventually, once the inter-commutings start and the network erases the
`memory' of its initial conditions,
this velocity difference is also gradually erased.

Finally, we also notice that the network keeps chopping off loops
throughout the simulation, and that there is a dramatic increase
in the small scale structure of the network, particularly at later times.
Visually, the string network develops large  numbers of 
`knots', highly convoluted strings regions where the wiggly long strings 
have collapsed inhomogeneously.
These small scale features have proved to be difficult to evolve numerically,
and this in fact turns out to be the main limiting factor
which at present preventing us
from running the simulations closer to the big crunch.

\subsection{Contrasting the analytical and numerical approaches}

\begin{figure}
\includegraphics[width=3in,keepaspectratio]{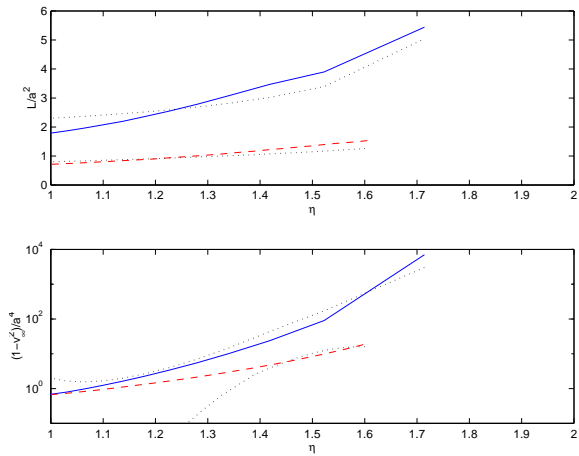}
\caption{\label{tests}
Comparing our analytic solutions for the behaviour of cosmic
string networks as they approach the asymptotic regime
with numerical simulations in the matter
(solid) and radiation (dashed) epochs. The top panel refers to
our solutions for the correlation length (\ref{asymprad}-\ref{asympmat}),
while the bottom panel refers to the solutions for the
long string RMS velocity (\ref{asympvvrad}-\ref{asympvvmat}).
In all cases the dotted curves are obtained using the analytic
solutions and a loop chopping efficiency ${\tilde c}=0.23$.}
\end{figure}

We end this section by testing our solutions for the evolution of
the string network in the contracting phase, obtained with the
VOS model, against our numerical simulations. The outcome
of these tests is summarized in Fig. \ref{tests}.

We compare our solutions (\ref{asymprad}-\ref{asympmat}) for the
long string correlation length and
(\ref{asympvvrad}-\ref{asympvvmat}) for the long string RMS velocity with
the numerical simulations described above, by plotting respectively
$L/a^2$ and $(1-v^2)/a^4$. We have assumed the usual \textit{ansatz}
for the momentum parameter $k(v)$---see Eqn. (\ref{kkvvv})
and \cite{Martins3}---while for the
loop chopping efficiency we have assumed a constant value ${\tilde c}=0.23$
which was obtained in previous studies of high resolution simulations
in expanding universes \cite{Martins3,Moore}.

We can see that, given the approximate nature of our solutions and
the numerical errors in the simulations, the matching between the
two is fairly remarkable. Naturally the quality of the fit will be
crucially determined by our assumptions about ${\tilde c}$
and $k(v)$, and as such one could consider this an independent test
on the behaviour of these parameters. However, given the uncertainties
discussed above, one can not really meaningfully use these simulations
to `measure' ${\tilde c}$ directly from the simulations (with robust
error bars) and test the $k(v)$ \textit{ansatz}.

Among other reasons this is pertinent because, as mentioned above, there
could be further corrections to the equations in the analytic model.

We note that tentative measurements of the loop chopping efficiency
${\tilde c}$ directly from the simulations (that is, without using any
VOS model dependent assumptions) are consistent with no variation throughout
the range probed by the simulations, though again
error bars in this particular measurement are significant. (This is
because this measurement requires the calculation
of second derivatives of quantities in the simulation, whereas all
quantities plotted in this paper require, at most, the calculation of
first derivatives, and are therefore much more stable numerically). 
However, the limited dynamical range of the simulation does 
not allow us to distinguish between the two discussed \textit{ansatze} for 
the evolution of ${\tilde c}$ (namely ${\tilde c} =  constant$ and 
${\tilde c} \propto \gamma_\infty^{-1/2}$).

If anything, one could use the observation
that in Fig. \ref{tests} the numerical curves are
steeper than the analytic ones to infer
that our approximate solution is \textit{underestimating} the loop
production, though on the other hand this could be due to further
corrections on $k(v)$ rather than to corrections on ${\tilde c}$.
In any case, it is clear from direct inspection of snapshots of the
simulation boxes that the amount of small-scale wiggles is gradually
building up.

Be that as it may, the two extreme cases outlined above are likely to
be applicable only in asymptotic regimes.  
The loop chopping efficiency is clearly important initially but it cannot
continue to grow
arbitrarily large, since the amount of energy transferred to
loops (and small scale wiggles) at any time is limited by
causality. On the other hand, a scenario in which loop production switches
off completely may be attained ultimately during the final collapse. 
However, this is clearly
delayed by the obvious build-up of
small-scale structure on the strings.  
So at the moment, given the finite resolution of the simulations,
the quantitative behaviour of the network is somewhat open to debate. 
Any more accurate modelling can only be meaningfully done
in the context of a proper multi-scale string evolution model
which, although possible \cite{Austin,Thesis,Wiggly} is beyond the scope of
the present work.

Finally it is also worth keeping in mind that any discussion of the
evolution of a cosmic string network with the present formalism is only
applicable while one is well below the Hagedorn temperature, at which 
the strings would `dissolve' in a reverse phase transition.
Discussions
of asymptotic regimes should be taken with some caution, since 
a cosmic string network will only survive the bounce intact if this happens
before the Hagedorn temperature is reached.

\section{\label{conseq}Discussion: cosmological consequences}

The evolution of the string energy density is dependent on the 
dynamics of the universe. In an expanding universe the long string energy 
density will evolve as 
\be\label{density}
\rho_\infty \propto a^n
\ee
where $n\approx -4$ during the radiation-dominated era, $n\approx -3$
during the matter-dominated era and $n\approx -2$ during a
curvature-dominated or accelerated expansion era.  The overall density
of strings remains constant relative to the background
density $\bar\rho$ in both radiation and matter eras
\be
\frac{\rho_\infty}{\bar\rho} = \sigma G\mu\,,
\ee
with $\sigma _{\rm r}\approx 400$ and $\sigma_{\rm m}\approx 60$ respectively
\cite{Bennett,Allen1}.
During curvature domination or accelerated expansion, the string density grows
relative to the other matter as 
$\rho_\infty/\rho_{\rm m} \propto a$.   For GUT-scale strings 
with $G\mu \sim 10^{-6}$ this gives the interesting conclusion that today 
strings have a comparable energy density to the cosmic microwave background
radiation $\rho_{\rm cmb}/\rho_{\rm m}\sim 10^{-4}$.  However, a realistic 
cyclic model will continue
to expand well beyond $t_0$, so the string density at maximum expansion 
will end up being much greater than the radiation density (even for strings
considerably lighter than GUT scale). 
In addition, the gravitational (or other) radiation produced through 
the continuous decay of the string network
evolves as 
\be\label{grdensity}
\rho_{\rm gr} \propto a^{-4}\,.
\ee
It might appear that this contribution would become negligible during the 
matter era but in each Hubble time the strings lose about half their energy 
into gravitational radiation, so this background always remains comparable
to the string density $\rho_{\rm gr}\sim \rho_\infty$.

\subsection{A string-dominated universe}

Now consider the collapsing phase in which the string network, like
the gravitational waves they have produced, begins to behave like radiation.  
Globally, the density of both the strings and the gravitational waves 
will grow as $a^{-4}$ and, together with any other radiation components,
they will eventually dominate over any nonrelativistic matter.  
In a realistic cyclic model reaching maximum expansion in the far 
distant future, sufficiently massive strings and their decay products 
will have a greater density than the cosmic microwave and neutrino backgrounds.
As the universe contracts, then, it will eventually reach a state in 
which the relativistic string network and/or their gravitational waves
dominate the global dynamics of the universe!  This seems likely to lead to 
a dramatically 
different energy content for the universe after it emerges from the 
next bounce.

At this stage we should note, however, that if the gravitational radiation
background becomes dominated by the longer wavelength modes then the radiation
fluid approximation will eventually break down, and one then expects
that this background will behave as `curvature'
rather than radiation---this is shown, in an the context of inflation,
in \cite{Abramo}. Obviously perturbations in such modes can't be directly
detected or have a direct impact on cosmological observables, but they
can have an impact on the background in which they propagate.
On the other hand, it is not yet clear if these results are directly applicable
to the contracting case, and if and when that regime is reached, since in
order for the long wavelength modes to become dominant one requires that
loop production (and hence gravitational radiation) switches off fast
enough on small (sub-horizon) scales.

Even for lighter strings which do not dominate the universe,  
they would end up with a much greater density in the collapsing phase 
than they had previously during expansion.
If the universe went through a bounce, the energy density
in the cosmic strings and gravitational radiation produced by the network 
would be much greater after the bounce than before it (though the exact amount is 
dependent on the model details, in particular on the duration of the matter
and curvature and/or accelerated expansion era).
For example, bounds on the string mass per unit length obtained in order not to
overproduce a gravitational radiation background may be severely modified,
in addition to the more general constraints on additional relativistic
fluids \cite{Burles,Lisi,Bowen}.

\subsection{An inhomogeneous universe}

Furthermore, unlike the uniform CMB background, 
the energy density in both cosmic strings and gravitational
radiation will be very inhomogeneous.  
In the collapsing regime,
we expect that an increasingly small fraction of Hubble regions will have a string
passing through them. Those that do will become string dominated since the 
string energy density in those regions will approximately evolve as
\be\label{regions}
\frac{\rho_\infty}{\bar \rho} \propto \gamma_\infty \propto a^{-2}\,,
\ee
up to the corrections described in the previous section. This means 
that for these regions
the assumption of a FRW background will cease to be valid at sufficiently
late times, and the defects can make the universe
very inhomogeneous \cite{Inhomog} or even anisotropic \cite{Fossils}.
Even Hubble regions without strings can be expected to have large fluctuations
in their gravitational radiation content.  For sufficiently 
massive strings, both of these effects
can be expected to survive the bounce to create large inhomogeneities 
in the next cycle.

We conclude that a cosmic string network will be a significant source of
radiation, entropy and inhomogeneity
which may be problematic in the cyclic context. 
An attempt to solve the problem of the overproduction of entropy and
unwanted relics has been proposed in the ekpyrotic context \cite{Steinhardt}.
They suggest a cyclic model for the universe where
an extended period of cosmic acceleration at low energies is used
to remove the entropy, black holes, and other debris produced in the
previous cycle. It is clear that when quantitative
calculations are carried out  to
establish the amount of acceleration required to
dilute unwanted debris, then the answer will depend on whether or
not cosmic strings or other topological defects are present; 
a much longer period could
be required if they are.

We note that something analogous happens for the case of black
holes---see \cite{Barrow} for a discussion of this case. From cycle
to cycle one expects that they will accumulate, since they will
tend to have Hawking lifetimes longer than the duration of the cycle.
The only ways to get rid of them are having a bouncing universe
that is very close to flatness (so as to increase the duration of
the cycle) or having then annihilated or torn apart at the bounce
singularity.

Finally, some of the results described in this paper are also expected to
be valid for other topological defects, in particular domain
walls \cite{Book}. We shall reserve a more detailed analysis of the 
evolution and implications of strings and other defects in the context 
of bouncing universes to a forthcoming publication.

\section{\label{conc}Conclusions}

In this paper we have presented a first study of the basic evolutionary
properties of cosmic string networks in contracting universes, using
both analytic methods and high-resolution numerical simulations.
We have shown that the string network becomes ultra-relativistic,
and at late times will approximately behave like a radiation fluid.
We have derived new analytic scaling solutions describing this
behaviour and shown, through high-resolution numerical simulations,
that these analytic solutions are a good approximation to the
actual string dynamics.

The main cosmological consequence of this asymmetric behaviour in the
evolution of cosmic string networks in the collapsing and expanding phases
is that it makes them a significant source of entropy and inhomogeneity, 
and therefore establishes the need for a suitable entropy dilution mechanism 
if they
are present in a bouncing cosmological scenario. This mechanism
will also operate, \textit{mutatis mutandis}, for other stable
topological defects. Conversely, if
direct evidence is found for the presence of topological defects
(with a given energy scale) in the early universe, their existence
alone will impose constraints on the existence and characteristics of
any previous phases of cosmological collapse.

\begin{acknowledgments}
We thank John Barrow, Kostas Dimopoulos, Ruth Durrer, Ruth Gregory and
David Wands for useful discussions and
comments at various stages of this work.  The string evolution code
used for the numerical simulations was developed by E.P.S. in collaboration 
with Bruce Allen.
C.S. thanks the Mathematical Sciences Institute in Durham
for hospitality during part of this work.  

C.M. and C.S. are funded by FCT (Portugal), under grants nos.
FMRH/BPD/1600/2000 and BPD/22092/99 respectively.
Additional support for this project came from grant CERN/FIS/43737/2001.

This work was done in the context of the COSLAB network, and
was performed on COSMOS, the Origin3800 owned by the UK
Computational Cosmology Consortium, supported by SGI, HEFCE and PPARC.
\end{acknowledgments}

\bibliography{cyclic}

\begin{thebibliography}{54}
\expandafter\ifx\csname natexlab\endcsname\relax\def\natexlab#1{#1}\fi
\expandafter\ifx\csname bibnamefont\endcsname\relax
  \def\bibnamefont#1{#1}\fi
\expandafter\ifx\csname bibfnamefont\endcsname\relax
  \def\bibfnamefont#1{#1}\fi
\expandafter\ifx\csname citenamefont\endcsname\relax
  \def\citenamefont#1{#1}\fi
\expandafter\ifx\csname url\endcsname\relax
  \def\url#1{\texttt{#1}}\fi
\expandafter\ifx\csname urlprefix\endcsname\relax\def\urlprefix{URL }\fi
\providecommand{\bibinfo}[2]{#2}
\providecommand{\eprint}[2][]{\url{#2}}

\bibitem[{\citenamefont{Tolman}(1934)}]{Tolman}
\bibinfo{author}{\bibfnamefont{R.~C.} \bibnamefont{Tolman}}
  (\bibinfo{year}{1934}), \bibinfo{note}{{ }Oxford, U.K.: Clarendon Press}.

\bibitem[{\citenamefont{Steinhardt and Turok}(2002)}]{Steinhardt}
\bibinfo{author}{\bibfnamefont{P.~J.} \bibnamefont{Steinhardt}}
  \bibnamefont{and} \bibinfo{author}{\bibfnamefont{N.}~\bibnamefont{Turok}},
  \bibinfo{journal}{Phys. Rev.} \textbf{\bibinfo{volume}{D65}},
  \bibinfo{pages}{126003} (\bibinfo{year}{2002}), \eprint{hep-th/0111098}.

\bibitem[{\citenamefont{Kanekar et~al.}(2001)\citenamefont{Kanekar, Sahni, and
  Shtanov}}]{Kanekar}
\bibinfo{author}{\bibfnamefont{N.}~\bibnamefont{Kanekar}},
  \bibinfo{author}{\bibfnamefont{V.}~\bibnamefont{Sahni}}, \bibnamefont{and}
  \bibinfo{author}{\bibfnamefont{Y.}~\bibnamefont{Shtanov}},
  \bibinfo{journal}{Phys. Rev.} \textbf{\bibinfo{volume}{D63}},
  \bibinfo{pages}{083520} (\bibinfo{year}{2001}),
  \eprint[http://arXiv.org/abs]{astro-ph/0101448}.

\bibitem[{\citenamefont{Peter and Pinto-Neto}(2002{\natexlab{a}})}]{Peter1}
\bibinfo{author}{\bibfnamefont{P.}~\bibnamefont{Peter}} \bibnamefont{and}
  \bibinfo{author}{\bibfnamefont{N.}~\bibnamefont{Pinto-Neto}},
  \bibinfo{journal}{Phys. Rev.} \textbf{\bibinfo{volume}{D65}},
  \bibinfo{pages}{023513} (\bibinfo{year}{2002}{\natexlab{a}}),
  \eprint[http://arXiv.org/abs]{gr-qc/0109038}.

\bibitem[{\citenamefont{Peter and Pinto-Neto}(2002{\natexlab{b}})}]{Peter2}
\bibinfo{author}{\bibfnamefont{P.}~\bibnamefont{Peter}} \bibnamefont{and}
  \bibinfo{author}{\bibfnamefont{N.}~\bibnamefont{Pinto-Neto}},
  \bibinfo{journal}{Phys. Rev.} \textbf{\bibinfo{volume}{D66}},
  \bibinfo{pages}{063509} (\bibinfo{year}{2002}{\natexlab{b}}),
  \eprint{hep-th/0203013}.

\bibitem[{\citenamefont{Kibble}(1976)}]{Kibble1}
\bibinfo{author}{\bibfnamefont{T.~W.~B.} \bibnamefont{Kibble}},
  \bibinfo{journal}{J. Phys.} \textbf{\bibinfo{volume}{A9}},
  \bibinfo{pages}{1387} (\bibinfo{year}{1976}).

\bibitem[{\citenamefont{Vilenkin and Shellard}(1994)}]{Book}
\bibinfo{author}{\bibfnamefont{A.}~\bibnamefont{Vilenkin}} \bibnamefont{and}
  \bibinfo{author}{\bibfnamefont{E.~P.~S.} \bibnamefont{Shellard}}
  (\bibinfo{year}{1994}), \bibinfo{note}{{ }Cambridge, U.K.: Cambridge
  University Press}.

\bibitem[{\citenamefont{Avelino et~al.}(2002)\citenamefont{Avelino, Martins,
  Santos, and Shellard}}]{Contracting}
\bibinfo{author}{\bibfnamefont{P.~P.} \bibnamefont{Avelino}},
  \bibinfo{author}{\bibfnamefont{C.~J. A.~P.} \bibnamefont{Martins}},
  \bibinfo{author}{\bibfnamefont{C.}~\bibnamefont{Santos}}, \bibnamefont{and}
  \bibinfo{author}{\bibfnamefont{E.~P.~S.} \bibnamefont{Shellard}},
  \bibinfo{journal}{Phys. Rev. Lett.} \textbf{\bibinfo{volume}{89}},
  \bibinfo{pages}{271301} (\bibinfo{year}{2002}), \eprint{astro-ph/0211066}.

\bibitem[{\citenamefont{Kibble}(1985)}]{Kibble}
\bibinfo{author}{\bibfnamefont{T.~W.~B.} \bibnamefont{Kibble}},
  \bibinfo{journal}{Nucl. Phys.} \textbf{\bibinfo{volume}{B252}},
  \bibinfo{pages}{227} (\bibinfo{year}{1985}).

\bibitem[{\citenamefont{Austin et~al.}(1993)\citenamefont{Austin, Copeland, and
  Kibble}}]{Austin}
\bibinfo{author}{\bibfnamefont{D.}~\bibnamefont{Austin}},
  \bibinfo{author}{\bibfnamefont{E.~J.} \bibnamefont{Copeland}},
  \bibnamefont{and} \bibinfo{author}{\bibfnamefont{T.~W.~B.}
  \bibnamefont{Kibble}}, \bibinfo{journal}{Phys. Rev.}
  \textbf{\bibinfo{volume}{D48}}, \bibinfo{pages}{5594} (\bibinfo{year}{1993}),
  \eprint[http://arXiv.org/abs]{hep-ph/9307325}.

\bibitem[{\citenamefont{Martins and Shellard}(1996{\natexlab{a}})}]{Martins1}
\bibinfo{author}{\bibfnamefont{C.~J. A.~P.} \bibnamefont{Martins}}
  \bibnamefont{and} \bibinfo{author}{\bibfnamefont{E.~P.~S.}
  \bibnamefont{Shellard}}, \bibinfo{journal}{Phys. Rev.}
  \textbf{\bibinfo{volume}{D53}}, \bibinfo{pages}{575}
  (\bibinfo{year}{1996}{\natexlab{a}}),
  \eprint[http://arXiv.org/abs]{hep-ph/9507335}.

\bibitem[{\citenamefont{Martins}(1997{\natexlab{a}})}]{Thesis}
\bibinfo{author}{\bibfnamefont{C.~J. A.~P.} \bibnamefont{Martins}}
  (\bibinfo{year}{1997}{\natexlab{a}}), \bibinfo{note}{{ }PhD Thesis,
  University of Cambridge}.

\bibitem[{\citenamefont{Martins and Shellard}(2002)}]{Martins3}
\bibinfo{author}{\bibfnamefont{C.~J. A.~P.} \bibnamefont{Martins}}
  \bibnamefont{and} \bibinfo{author}{\bibfnamefont{E.~P.~S.}
  \bibnamefont{Shellard}}, \bibinfo{journal}{Phys. Rev.}
  \textbf{\bibinfo{volume}{D65}}, \bibinfo{pages}{043514}
  (\bibinfo{year}{2002}), \eprint[http://arXiv.org/abs]{hep-ph/0003298}.

\bibitem[{\citenamefont{Bennett and Bouchet}(1989)}]{Bennett}
\bibinfo{author}{\bibfnamefont{D.~P.} \bibnamefont{Bennett}} \bibnamefont{and}
  \bibinfo{author}{\bibfnamefont{F.~R.} \bibnamefont{Bouchet}},
  \bibinfo{journal}{Phys. Rev. Lett.} \textbf{\bibinfo{volume}{63}},
  \bibinfo{pages}{2776} (\bibinfo{year}{1989}).

\bibitem[{\citenamefont{Allen and Shellard}(1990)}]{Allen1}
\bibinfo{author}{\bibfnamefont{B.}~\bibnamefont{Allen}} \bibnamefont{and}
  \bibinfo{author}{\bibfnamefont{E.~P.~S.} \bibnamefont{Shellard}},
  \bibinfo{journal}{Phys. Rev. Lett.} \textbf{\bibinfo{volume}{64}},
  \bibinfo{pages}{119} (\bibinfo{year}{1990}).

\bibitem[{\citenamefont{Moore et~al.}(2002)\citenamefont{Moore, Shellard, and
  Martins}}]{Moore}
\bibinfo{author}{\bibfnamefont{J.~N.} \bibnamefont{Moore}},
  \bibinfo{author}{\bibfnamefont{E.~P.~S.} \bibnamefont{Shellard}},
  \bibnamefont{and} \bibinfo{author}{\bibfnamefont{C.~J. A.~P.}
  \bibnamefont{Martins}}, \bibinfo{journal}{Phys. Rev.}
  \textbf{\bibinfo{volume}{D65}}, \bibinfo{pages}{023503}
  (\bibinfo{year}{2002}), \eprint[http://arXiv.org/abs]{hep-ph/0107171}.

\bibitem[{\citenamefont{Avelino et~al.}(1998)\citenamefont{Avelino, Shellard,
  Wu, and Allen}}]{Avelino2}
\bibinfo{author}{\bibfnamefont{P.~P.} \bibnamefont{Avelino}},
  \bibinfo{author}{\bibfnamefont{E.~P.~S.} \bibnamefont{Shellard}},
  \bibinfo{author}{\bibfnamefont{J.~H.~P.} \bibnamefont{Wu}}, \bibnamefont{and}
  \bibinfo{author}{\bibfnamefont{B.}~\bibnamefont{Allen}},
  \bibinfo{journal}{Phys. Rev. Lett.} \textbf{\bibinfo{volume}{81}},
  \bibinfo{pages}{2008} (\bibinfo{year}{1998}),
  \eprint[http://arXiv.org/abs]{astro-ph/9712008}.

\bibitem[{\citenamefont{Avelino and Martins}(2000{\natexlab{a}})}]{Adiabatic}
\bibinfo{author}{\bibfnamefont{P.~P.} \bibnamefont{Avelino}} \bibnamefont{and}
  \bibinfo{author}{\bibfnamefont{C.~J. A.~P.} \bibnamefont{Martins}},
  \bibinfo{journal}{Phys. Rev. Lett.} \textbf{\bibinfo{volume}{85}},
  \bibinfo{pages}{1370} (\bibinfo{year}{2000}{\natexlab{a}}),
  \eprint[http://arXiv.org/abs]{astro-ph/0002413}.

\bibitem[{\citenamefont{Avelino and Martins}(2001)}]{Gaussian}
\bibinfo{author}{\bibfnamefont{P.~P.} \bibnamefont{Avelino}} \bibnamefont{and}
  \bibinfo{author}{\bibfnamefont{C.~J. A.~P.} \bibnamefont{Martins}},
  \bibinfo{journal}{Phys. Lett.} \textbf{\bibinfo{volume}{B516}},
  \bibinfo{pages}{191} (\bibinfo{year}{2001}),
  \eprint[http://arXiv.org/abs]{astro-ph/0006303}.

\bibitem[{\citenamefont{Avelino and Martins}(2000{\natexlab{b}})}]{Fossils}
\bibinfo{author}{\bibfnamefont{P.~P.} \bibnamefont{Avelino}} \bibnamefont{and}
  \bibinfo{author}{\bibfnamefont{C.~J. A.~P.} \bibnamefont{Martins}},
  \bibinfo{journal}{Phys. Rev.} \textbf{\bibinfo{volume}{D62}},
  \bibinfo{pages}{103510} (\bibinfo{year}{2000}{\natexlab{b}}),
  \eprint[http://arXiv.org/abs]{astro-ph/0003231}.

\bibitem[{\citenamefont{Avelino et~al.}(2001)\citenamefont{Avelino, Carvalho,
  Martins, and Oliveira}}]{Inhomog}
\bibinfo{author}{\bibfnamefont{P.~P.} \bibnamefont{Avelino}},
  \bibinfo{author}{\bibfnamefont{J.~P. M.~d.} \bibnamefont{Carvalho}},
  \bibinfo{author}{\bibfnamefont{C.~J. A.~P.} \bibnamefont{Martins}},
  \bibnamefont{and} \bibinfo{author}{\bibfnamefont{J.~C. R.~E.}
  \bibnamefont{Oliveira}}, \bibinfo{journal}{Phys. Lett.}
  \textbf{\bibinfo{volume}{B515}}, \bibinfo{pages}{148} (\bibinfo{year}{2001}),
  \eprint[http://arXiv.org/abs]{astro-ph/0004227}.

\bibitem[{\citenamefont{Rees}(1969)}]{Rees}
\bibinfo{author}{\bibfnamefont{M.~J.} \bibnamefont{Rees}},
  \bibinfo{journal}{The Observatory} \textbf{\bibinfo{volume}{89}},
  \bibinfo{pages}{193} (\bibinfo{year}{1969}).

\bibitem[{\citenamefont{Zeldovich and Novikov}(1983)}]{Zeldovich}
\bibinfo{author}{\bibfnamefont{Y.~B.} \bibnamefont{Zeldovich}}
  \bibnamefont{and} \bibinfo{author}{\bibfnamefont{I.~D.}
  \bibnamefont{Novikov}} (\bibinfo{year}{1983}), \bibinfo{note}{{ }Chicago,
  U.S.A.: University of Chicago Press}.

\bibitem[{\citenamefont{Markov}(1984)}]{Markov}
\bibinfo{author}{\bibfnamefont{M.~A.} \bibnamefont{Markov}},
  \bibinfo{journal}{Annals Phys.} \textbf{\bibinfo{volume}{155}},
  \bibinfo{pages}{333} (\bibinfo{year}{1984}).

\bibitem[{\citenamefont{Penrose}(1990)}]{Penrose}
\bibinfo{author}{\bibfnamefont{R.}~\bibnamefont{Penrose}},
  \bibinfo{journal}{Ann. N. Y. Acad. Sci.} \textbf{\bibinfo{volume}{571}},
  \bibinfo{pages}{249} (\bibinfo{year}{1990}).

\bibitem[{\citenamefont{Barrow and Dabrowski}(1995)}]{Barrow}
\bibinfo{author}{\bibfnamefont{J.}~\bibnamefont{Barrow}} \bibnamefont{and}
  \bibinfo{author}{\bibfnamefont{M.}~\bibnamefont{Dabrowski}},
  \bibinfo{journal}{M.N.R.A.S.} \textbf{\bibinfo{volume}{275}},
  \bibinfo{pages}{850} (\bibinfo{year}{1995}).

\bibitem[{\citenamefont{Dabrowski}(1996)}]{Dabrowski}
\bibinfo{author}{\bibfnamefont{M.~P.} \bibnamefont{Dabrowski}},
  \bibinfo{journal}{Ann. Phys.} \textbf{\bibinfo{volume}{248}},
  \bibinfo{pages}{199} (\bibinfo{year}{1996}).

\bibitem[{\citenamefont{Durrer and Laukenmann}(1996)}]{Durrer}
\bibinfo{author}{\bibfnamefont{R.}~\bibnamefont{Durrer}} \bibnamefont{and}
  \bibinfo{author}{\bibfnamefont{J.}~\bibnamefont{Laukenmann}},
  \bibinfo{journal}{Class. Quant. Grav.} \textbf{\bibinfo{volume}{13}},
  \bibinfo{pages}{1069} (\bibinfo{year}{1996}),
  \eprint[http://arXiv.org/abs]{gr-qc/9510041}.

\bibitem[{\citenamefont{Hawking}(1984)}]{Hawking}
\bibinfo{author}{\bibfnamefont{S.~W.} \bibnamefont{Hawking}},
  \bibinfo{journal}{Nucl. Phys.} \textbf{\bibinfo{volume}{B239}},
  \bibinfo{pages}{257} (\bibinfo{year}{1984}).

\bibitem[{\citenamefont{Page}(1984)}]{Page}
\bibinfo{author}{\bibfnamefont{D.~N.} \bibnamefont{Page}},
  \bibinfo{journal}{Class. Quant. Grav.} \textbf{\bibinfo{volume}{417}},
  \bibinfo{pages}{1} (\bibinfo{year}{1984}).

\bibitem[{\citenamefont{Laflamme and Shellard}(1987)}]{Laflamme}
\bibinfo{author}{\bibfnamefont{R.}~\bibnamefont{Laflamme}} \bibnamefont{and}
  \bibinfo{author}{\bibfnamefont{E.~P.~S.} \bibnamefont{Shellard}},
  \bibinfo{journal}{Phys. Rev.} \textbf{\bibinfo{volume}{D35}},
  \bibinfo{pages}{2315} (\bibinfo{year}{1987}).

\bibitem[{\citenamefont{Cornish and Shellard}(1998)}]{Cornish}
\bibinfo{author}{\bibfnamefont{N.}~\bibnamefont{Cornish}} \bibnamefont{and}
  \bibinfo{author}{\bibfnamefont{E.~P.~S.} \bibnamefont{Shellard}},
  \bibinfo{journal}{Phys. Rev. Lett.} \textbf{\bibinfo{volume}{81}},
  \bibinfo{pages}{3571} (\bibinfo{year}{1998}).

\bibitem[{\citenamefont{Khoury et~al.}(2001)\citenamefont{Khoury, Ovrut,
  Steinhardt, and Turok}}]{Steinhardtetal}
\bibinfo{author}{\bibfnamefont{J.}~\bibnamefont{Khoury}},
  \bibinfo{author}{\bibfnamefont{B.~A.} \bibnamefont{Ovrut}},
  \bibinfo{author}{\bibfnamefont{P.~J.} \bibnamefont{Steinhardt}},
  \bibnamefont{and} \bibinfo{author}{\bibfnamefont{N.}~\bibnamefont{Turok}},
  \bibinfo{journal}{Phys. Rev.} \textbf{\bibinfo{volume}{D64}},
  \bibinfo{pages}{123522} (\bibinfo{year}{2001}),
  \eprint[http://arXiv.org/abs]{hep-th/0103239}.

\bibitem[{\citenamefont{Khoury et~al.}(2002)\citenamefont{Khoury, Ovrut,
  Seiberg, Steinhardt, and Turok}}]{Steinhardtetal1}
\bibinfo{author}{\bibfnamefont{J.}~\bibnamefont{Khoury}},
  \bibinfo{author}{\bibfnamefont{B.~A.} \bibnamefont{Ovrut}},
  \bibinfo{author}{\bibfnamefont{N.}~\bibnamefont{Seiberg}},
  \bibinfo{author}{\bibfnamefont{P.~J.} \bibnamefont{Steinhardt}},
  \bibnamefont{and} \bibinfo{author}{\bibfnamefont{N.}~\bibnamefont{Turok}},
  \bibinfo{journal}{Phys. Rev.} \textbf{\bibinfo{volume}{D65}},
  \bibinfo{pages}{086007} (\bibinfo{year}{2002}),
  \eprint[http://arXiv.org/abs]{hep-th/0108187}.

\bibitem[{\citenamefont{Brandenberger and Finelli}(2001)}]{Finelli}
\bibinfo{author}{\bibfnamefont{R.}~\bibnamefont{Brandenberger}}
  \bibnamefont{and} \bibinfo{author}{\bibfnamefont{F.}~\bibnamefont{Finelli}},
  \bibinfo{journal}{JHEP} \textbf{\bibinfo{volume}{11}}, \bibinfo{pages}{056}
  (\bibinfo{year}{2001}), \eprint[http://arXiv.org/abs]{hep-th/0109004}.

\bibitem[{\citenamefont{Lyth}(2002)}]{Lyth}
\bibinfo{author}{\bibfnamefont{D.~H.} \bibnamefont{Lyth}},
  \bibinfo{journal}{Phys. Lett.} \textbf{\bibinfo{volume}{B526}},
  \bibinfo{pages}{173} (\bibinfo{year}{2002}),
  \eprint[http://arXiv.org/abs]{hep-ph/0110007}.

\bibitem[{\citenamefont{Martin et~al.}(2002)\citenamefont{Martin, Peter,
  Pinto~Neto, and Schwarz}}]{Martin1}
\bibinfo{author}{\bibfnamefont{J.}~\bibnamefont{Martin}},
  \bibinfo{author}{\bibfnamefont{P.}~\bibnamefont{Peter}},
  \bibinfo{author}{\bibfnamefont{N.}~\bibnamefont{Pinto~Neto}},
  \bibnamefont{and} \bibinfo{author}{\bibfnamefont{D.~J.}
  \bibnamefont{Schwarz}}, \bibinfo{journal}{Phys. Rev.}
  \textbf{\bibinfo{volume}{D65}}, \bibinfo{pages}{123513}
  (\bibinfo{year}{2002}), \eprint{hep-th/0112128}.

\bibitem[{\citenamefont{Martin et~al.}(2003)\citenamefont{Martin, Peter,
  Pinto-Neto, and Schwarz}}]{Martin2}
\bibinfo{author}{\bibfnamefont{J.}~\bibnamefont{Martin}},
  \bibinfo{author}{\bibfnamefont{P.}~\bibnamefont{Peter}},
  \bibinfo{author}{\bibfnamefont{N.}~\bibnamefont{Pinto-Neto}},
  \bibnamefont{and} \bibinfo{author}{\bibfnamefont{D.~J.}
  \bibnamefont{Schwarz}}, \bibinfo{journal}{Phys. Rev.}
  \textbf{\bibinfo{volume}{D67}}, \bibinfo{pages}{028301}
  (\bibinfo{year}{2003}), \eprint{hep-th/0204222}.

\bibitem[{\citenamefont{Martins and Shellard}(1996{\natexlab{b}})}]{Martins2}
\bibinfo{author}{\bibfnamefont{C.~J. A.~P.} \bibnamefont{Martins}}
  \bibnamefont{and} \bibinfo{author}{\bibfnamefont{E.~P.~S.}
  \bibnamefont{Shellard}}, \bibinfo{journal}{Phys. Rev.}
  \textbf{\bibinfo{volume}{D54}}, \bibinfo{pages}{2535}
  (\bibinfo{year}{1996}{\natexlab{b}}),
  \eprint[http://arXiv.org/abs]{hep-ph/9602271}.

\bibitem[{\citenamefont{Martins}(1997{\natexlab{b}})}]{Open}
\bibinfo{author}{\bibfnamefont{C.~J. A.~P.} \bibnamefont{Martins}},
  \bibinfo{journal}{Phys. Rev.} \textbf{\bibinfo{volume}{D55}},
  \bibinfo{pages}{5208} (\bibinfo{year}{1997}{\natexlab{b}}),
  \eprint[http://arXiv.org/abs]{astro-ph/9701055}.

\bibitem[{\citenamefont{Martins and Shellard}(1997)}]{Condmat}
\bibinfo{author}{\bibfnamefont{C.~J. A.~P.} \bibnamefont{Martins}}
  \bibnamefont{and} \bibinfo{author}{\bibfnamefont{E.~P.~S.}
  \bibnamefont{Shellard}}, \bibinfo{journal}{Phys. Rev.}
  \textbf{\bibinfo{volume}{B56}}, \bibinfo{pages}{10892}
  (\bibinfo{year}{1997}), \eprint[http://arXiv.org/abs]{cond-mat/9607093}.

\bibitem[{\citenamefont{Avelino et~al.}(1997)\citenamefont{Avelino, Caldwell,
  and Martins}}]{Avelino1}
\bibinfo{author}{\bibfnamefont{P.~P.} \bibnamefont{Avelino}},
  \bibinfo{author}{\bibfnamefont{R.~R.} \bibnamefont{Caldwell}},
  \bibnamefont{and} \bibinfo{author}{\bibfnamefont{C.~J. A.~P.}
  \bibnamefont{Martins}}, \bibinfo{journal}{Phys. Rev.}
  \textbf{\bibinfo{volume}{D56}}, \bibinfo{pages}{4568} (\bibinfo{year}{1997}),
  \eprint[http://arXiv.org/abs]{astro-ph/9708057}.

\bibitem[{\citenamefont{Martins}(2000)}]{Wiggly}
\bibinfo{author}{\bibfnamefont{C.~J. A.~P.} \bibnamefont{Martins}},
  \bibinfo{journal}{Nucl. Phys. Proc. Suppl.} \textbf{\bibinfo{volume}{81}},
  \bibinfo{pages}{361} (\bibinfo{year}{2000}).

\bibitem[{\citenamefont{Carter}(1997)}]{School}
\bibinfo{author}{\bibfnamefont{B.}~\bibnamefont{Carter}}
  (\bibinfo{year}{1997}), \eprint[http://arXiv.org/abs]{hep-th/9705172}.

\bibitem[{\citenamefont{Vilenkin}(1991)}]{vilfric}
\bibinfo{author}{\bibfnamefont{A.}~\bibnamefont{Vilenkin}},
  \bibinfo{journal}{Phys. Rev.} \textbf{\bibinfo{volume}{D43}},
  \bibinfo{pages}{1060} (\bibinfo{year}{1991}).

\bibitem[{\citenamefont{Garfinkle and Will}(1987)}]{Garfinkle}
\bibinfo{author}{\bibfnamefont{D.}~\bibnamefont{Garfinkle}} \bibnamefont{and}
  \bibinfo{author}{\bibfnamefont{C.~M.} \bibnamefont{Will}},
  \bibinfo{journal}{Phys. Rev.} \textbf{\bibinfo{volume}{D35}},
  \bibinfo{pages}{1124} (\bibinfo{year}{1987}).

\bibitem[{\citenamefont{Avelino and Shellard}(1995)}]{Dynamical}
\bibinfo{author}{\bibfnamefont{P.~P.} \bibnamefont{Avelino}} \bibnamefont{and}
  \bibinfo{author}{\bibfnamefont{E.~P.~S.} \bibnamefont{Shellard}},
  \bibinfo{journal}{Phys. Rev.} \textbf{\bibinfo{volume}{D51}},
  \bibinfo{pages}{5946} (\bibinfo{year}{1995}).

\bibitem[{\citenamefont{Vilenkin}(1984)}]{DomV}
\bibinfo{author}{\bibfnamefont{A.}~\bibnamefont{Vilenkin}},
  \bibinfo{journal}{Phys. Rev. Lett.} \textbf{\bibinfo{volume}{53}},
  \bibinfo{pages}{1016} (\bibinfo{year}{1984}).

\bibitem[{\citenamefont{Kibble}(1986)}]{DomK}
\bibinfo{author}{\bibfnamefont{T.~W.~B.} \bibnamefont{Kibble}},
  \bibinfo{journal}{Phys. Rev.} \textbf{\bibinfo{volume}{D33}},
  \bibinfo{pages}{328} (\bibinfo{year}{1986}).

\bibitem[{\citenamefont{Kolb and Turner}(1990)}]{KTurner}
\bibinfo{author}{\bibfnamefont{E.~W.} \bibnamefont{Kolb}} \bibnamefont{and}
  \bibinfo{author}{\bibfnamefont{M.~S.} \bibnamefont{Turner}}
  (\bibinfo{year}{1990}), \bibinfo{note}{{ }Redwood City, U.S.A.:
  Addison-Wesley}.

\bibitem[{\citenamefont{Abramo}(1999)}]{Abramo}
\bibinfo{author}{\bibfnamefont{L.~R.~W.} \bibnamefont{Abramo}},
  \bibinfo{journal}{Phys. Rev.} \textbf{\bibinfo{volume}{D60}},
  \bibinfo{pages}{064004} (\bibinfo{year}{1999}),
  \eprint[http://arXiv.org/abs]{astro-ph/9903270}.

\bibitem[{\citenamefont{Burles et~al.}(1999)\citenamefont{Burles, Nollett,
  Truran, and Turner}}]{Burles}
\bibinfo{author}{\bibfnamefont{S.}~\bibnamefont{Burles}},
  \bibinfo{author}{\bibfnamefont{K.~M.} \bibnamefont{Nollett}},
  \bibinfo{author}{\bibfnamefont{J.~N.} \bibnamefont{Truran}},
  \bibnamefont{and} \bibinfo{author}{\bibfnamefont{M.~S.}
  \bibnamefont{Turner}}, \bibinfo{journal}{Phys. Rev. Lett.}
  \textbf{\bibinfo{volume}{82}}, \bibinfo{pages}{4176} (\bibinfo{year}{1999}),
  \eprint[http://arXiv.org/abs]{astro-ph/9901157}.

\bibitem[{\citenamefont{Lisi et~al.}(1999)\citenamefont{Lisi, Sarkar, and
  Villante}}]{Lisi}
\bibinfo{author}{\bibfnamefont{E.}~\bibnamefont{Lisi}},
  \bibinfo{author}{\bibfnamefont{S.}~\bibnamefont{Sarkar}}, \bibnamefont{and}
  \bibinfo{author}{\bibfnamefont{F.~L.} \bibnamefont{Villante}},
  \bibinfo{journal}{Phys. Rev.} \textbf{\bibinfo{volume}{D59}},
  \bibinfo{pages}{123520} (\bibinfo{year}{1999}),
  \eprint[http://arXiv.org/abs]{hep-ph/9901404}.

\bibitem[{\citenamefont{Bowen et~al.}(2002)\citenamefont{Bowen, Hansen,
  Melchiorri, Silk, and Trotta}}]{Bowen}
\bibinfo{author}{\bibfnamefont{R.}~\bibnamefont{Bowen}},
  \bibinfo{author}{\bibfnamefont{S.~H.} \bibnamefont{Hansen}},
  \bibinfo{author}{\bibfnamefont{A.}~\bibnamefont{Melchiorri}},
  \bibinfo{author}{\bibfnamefont{J.}~\bibnamefont{Silk}}, \bibnamefont{and}
  \bibinfo{author}{\bibfnamefont{R.}~\bibnamefont{Trotta}},
  \bibinfo{journal}{Mon. Not. Roy. Astron. Soc.}
  \textbf{\bibinfo{volume}{334}}, \bibinfo{pages}{760} (\bibinfo{year}{2002}),
  \eprint{astro-ph/0110636}.

\end{thebibliography}

\end{document}